\documentclass{mn2e}
\usepackage{graphicx}
\usepackage{subfigure}

\newcommand{\mnref}[1]{\hangindent=0.5in \hangafter=1 #1 \par}
\newenvironment{refs}{\parindent=0pt}{\parindent=1.5em}

\newcommand{\mnras}{MNRAS}

\newcommand{\aj}{AJ}
\newcommand{\apj}{ApJ}
\newcommand{\apjs}{ApJS}
\newcommand{\aaa}{A\&A}
\newcommand{\aap}{A\&A}
\newcommand{\aas}{A\&AS}

\newcommand{\Msolar}{\mbox{\,$\rm M_{\odot}$}}
\newcommand{\Lsolar}{\mbox{\,$\rm L_{\odot}$}}
\def\gs{\mathrel{\raise1.16pt\hbox{$>$}\kern-7.0pt
\lower3.06pt\hbox{{$\scriptstyle \sim$}}}}
\def\ls{\mathrel{\raise1.16pt\hbox{$<$}\kern-7.0pt
\lower3.06pt\hbox{{$\scriptstyle \sim$}}}}

\title{Tracers of Discs and Winds around Intermediate and High Mass Young Stellar Objects}
\author[S.L. Lumsden, H.E. Wheelwright, M.G. Hoare, R.D. Oudmaijer, J.E. Drew]
{S.L. Lumsden$^{1}$,  H.E. Wheelwright$^{1,2}$, M.G. Hoare$^{1}$
R.D. Oudmaijer$^{1}$ and J.E. Drew$^{3}$\\
{}$^1$ {\em Department of Physics and Astronomy,
University of Leeds, Leeds LS2 9JT, UK}\\
{}$^2$ {\em Max Planck Institute for Radio Astronomy, 53121 Bonn, Germany}\\
{}$^3$ {\em Department of Physical Sciences, University of Hertfordshire,
Hatfield, Hertfordshire, AL10 9AB, UK}\\
{Email -- s.l.lumsden@leeds.ac.uk}
}

\begin{document}

\label{firstpage}

\maketitle

\begin{abstract}
We present a study of the kinematical properties of a small sample of nearby
near-infrared bright massive and intermediate mass young stellar objects using
emission lines sensitive to discs and winds.  
We show for the first time that the broad ($\sim500$kms$^{-1}$) symmetric line
wings on the HI Brackett series lines are due to Stark broadening or electron
scattering, rather than pure Doppler broadening due to high speed motion.
The results are consistent 
with the presence of a very dense circumstellar environment.  In addition, many
of these lines show evidence for weak line self-absorption, suggestive of a
wind or disc-wind origin for that part of the absorbing material.  The weakness
of the self-absorption suggests a large opening angle for such an outflow.  We
also study the fluorescent 1.688$\mu$m FeII line, which is sensitive to
dense material.  We fitted a Keplerian disc model to this line, and
find reasonable fits in all bar one case, in agreement with previous finding
for classical Be stars that fluorescent iron transitions are reasonable disc tracers.
Overall the picture is one in which these stars still have accretion discs,
with a very dense inner circumstellar environment which may be tracing either
the inner regions of a disc, or of a stellar wind, and in which ionised outflow
is also present.  The similarity with lower mass stars is striking, suggesting
that at least in this mass range they form in a similar fashion.
\end{abstract}

\begin{keywords} accretion discs; line: profiles; stars: wind, outflows; stars:
  pre-main-sequence; 
  stars: formation
\end{keywords}

\section{Introduction}

There are well recognised and long-standing difficulties in describing
high-mass star formation as a simple scaled-up version of the low-mass case (eg
Kahn 1974, Shu, Adams \& Lizano 1987).  Radiation pressure is sufficient to
prevent further accretion once fusion commences in the stellar core, preventing
the formation of a star much more massive than 20$\Msolar$ in the simple low
mass theory.  This led to alternative models of high-mass star formation being
suggested (e.g.\ Bonnell, Vine \& Bate 2004, Stahler, Palla \& Ho 2000, McKee
\& Tan 2003) which can generally be subdivided into two groups: those that
overcome the mass limit by forming massive stars through competitive accretion
in dense cluster cores and those that look for a working version of the
scaled-up low-mass star formation paradigm, essentially overcoming the pressure
from the star through self-shielded disc accretion.  Dynamical models of disc
accretion around massive stars are now being produced which appear to overcome
the mass limit (eg Krumholz et al.\ 2009 who model the formation of a
40$\Msolar$ star, and Kuiper et al.\ 2010 who find radiation pressure is not an
efficient means of preventing disc accretion even in very massive stars).  Such
models are therefore currently favoured as the most likely.  This
agrees with limited observational data, where infrared interferometry has
revealed disk-like structures (eg.\ Kraus et al.\ 2010).
However we still have little direct evidence for accretion in a large sample of
young massive stars.

The obvious alternative method to apply to a large sample is to look for
kinematical evidence of a disc.  The 2.3$\mu$m CO first overtone
rotation-vibration emission is believed to mostly arise in an accretion disc in
young stars (eg, Carr 1989, Chandler, Carlstrom \& Scoville 1995, Bik \& Thi
2004, Blum et al.\ 2004, Wheelwright et al.\ 2010, Davies et al.\ 2010).  The
physical conditions required for its origin are characteristic of the
properties of a disc that is self-shielded (temperature in the range
1000--5000K, density greater than 10$^{10}$cm$^{-3}$), since it allows for the
presence of molecular gas relatively near the exciting star.  Generally,
the spectra of objects studied in this way have been better fitted by a disc
model than other alternatives.
However not all massive young stellar
objects show CO bandhead emission (eg Porter, Drew \& Lumsden 1998).  There are
many reasons why this might be, but the most plausible are that the accretion
rate may simply be too low to give detectable emission, or, there may be too
little column density to create the self-shielding condition required in the
disc to prevent the intense UV flux from more massive stars reaching the
mid-plane and destroying the CO.  The CO bandhead is also relatively complex to
interpret without high signal-to-noise data (eg Wheelwright et al) because of
the blending of the multiple rotational transitions that make up the $v=2-0$
vibrational transition.  

Another complicating factor in high mass stars is that other emission
mechanisms are also likely to be present.  The most obvious of these is a
stellar wind which can clearly be seen in at least some massive young stellar
objects (eg Drew, Bunn \& Hoare 1993, Bunn, Drew \& Hoare 1995, the models
presented by Sim, Drew \& Long 2005, and see Section 3.2).  A powerful stellar
wind is a fairly natural consequence of a high accretion rate in any hot star
system.  Winds in young stars are usually traced by recombination lines of
common elements, particularly hydrogen and helium (eg Simon et al.\ 1983 and
Persson et al.\ 1984 at moderate spectral resolution, and Bunn, Drew and Hoare
1995 at high enough resolution to fully sample the line profiles).
These lines can also be
excited in nebular gas and in a disc as well.  An ideal study should combine
tracers sensitive to the winds as well as those that sample density and
temperature and ionisation ranges more typical of an accretion disc in order to
clearly distinguish the origin of the emission.  Therefore we need a transition
that is only strongly excited at the higher densities seen in circumstellar
discs around young stellar objects (eg $n\gg 10^6$cm$^{-3}$).

Fortunately, there are other similar well studied situations from which we can
draw likely candidates.  For example, spectra of both active galactic nuclei
(eg Baldwin et al 2004) and classical Be stars (Zorec et al.\ 2007) show
fluorescent Fe$^+$ emission consistent with an origin in a disc.  It is not
seen in ordinary nebular gas (eg as in an HII region: see, for example, Lumsden
\& Puxley 1996).  FeII emission straddles the boundary between where CO
emission might still arise (neutral material) and ionised material since its
ionisation potential is only 7.9eV, with Fe$^{2+}$ having an ionisation
potential of 16.2eV.  Carciofi \& Bjorkman (2006) have successfully modelled
the origin of this emission in Be star disks from an origin in the shielded
material near the disk mid-plane.  Accretion discs around young stars are
likely to have additional opacity due to dust, but a similar partially ionised
zone will exist within them.

Fluorescent excitation typically occurs when an atom/ion is irradiated by
either UV continuum or Lyman series photons.  The excited atom/ion cascades
back down to the ground state, leading to transitions that would not typically
be seen from the standard recombination process.  
The optical FeII lines have been well studied and modelled in both active
galaxies and Be stars (eg Baldwin et al.\ 2004; Carciofi \& Bjorkman 2006) but
there are also infrared transitions that are more suitable for studying heavily
embedded young stellar objects.  In particular there are two strong lines near
1.6$\mu$m (eg see the spectrum of $\eta$ Carina in Hamann et al.\ 1994), of
which the transition at 1.6877808$\mu$m is both in a region of good atmospheric
transmission, and near enough in wavelength to the HI Br11 line that both can
be compared without the need to worry about differential extinction.  The
1.6877808$\mu$m c$^4$F$_9$--z$^4$F$_9$ transition arises from an upper level
6.2eV above the ground state (Johansson 1978) and is likely to be pumped by
Ly$\alpha$ in situations just such as we are seeking (the dense ionised/neutral
boundary) as discussed by Johansson \& Letokhov (2007).  Previous observations
(Porter et al.\ 1998) indicate this particular line may be relatively common in
massive young stellar objects, and more prevalent there than in lower mass
objects (Hamann \& Persson 1992).

The aim of this paper therefore is straightforward. We have observed a small
sample of eight intermediate and high mass young stellar objects with
sufficient spectral resolution to resolve the kinematics of the line emission
in both FeII and HI recombination lines for the first time.  The bulk of this
sample have near and/or mid-infrared interferometric data available, allowing
plausible limits to be placed on the inclination of the system and the size of
any infrared emitting region in a disc.  The comparison of both FeII and HI
tracers is ideal for testing whether the FeII samples a different emission
region from the HI recombination lines.  The sample is therefore ideal for
testing the value of the FeII line as a disc tracer, as well as expanding on
the use of HI lines as studied previously (eg Bunn et al 1995) to those present
in the $H$ band.  Our work presents a velocity resolved study of the higher
Brackett series members 11--4 and 12--4 for the first time.

In section 2 we discuss the sample and the observations made, in Section 3 we
discuss the observational results together with some simple model fits to our
data and in Section 4 we sum up the evidence as to which features might be the
best kinematical tracers of discs or winds around young massive stars.

\begin{table*}
\begin{tabular}{lrrrrrr}
Object & Distance (kpc) & Spectral type &Luminosity(\Lsolar) & $A_V$ (mag) & Inclination & $v_{LSR}$ (kms$^{-1}$)\\
BD+40$^\circ$4124/V1685 Cyg & $\sim$1$^{1\phantom{1}}$ & B3V$^{2\phantom{1}}$ && 3.1$^{3\phantom{1}}$ & $40-50^\circ$$^{4\phantom{1}}$  & $\sim$8$^{5\phantom{1}}$ \\
GL~490 & $\sim$1$^{6\phantom{1}}$ && 2000$^{6\phantom{1}}$ & 42$^{7\phantom{1}}$ & $\sim30^\circ$$^{8\phantom{1}}$ & $-$14$^{9\phantom{1}}$ \\
M17SW IRS1 & 1.3/2.1$^{14}$ & O9V$^{15}$ &&17$^{16}$ & $76^\circ$$^{15}$ & 
$\sim$17$^{17}$ \\
M8E & 1.25$^{10}$ && 20000$^{11}$ & 39$^{8\phantom{1}}$ & $\sim20^\circ$$^{12}$ & $
\sim11$$^{13}$ \\
MWC~297 & 0.25$^{18}$ & B1.5V$^{18}$ && 8$^{18}$ & 5$^\circ$$^{19}$, $<40^\circ$$^{20}$, $15^\circ$$^{21}$ & $2-20^{18}$ \\
MWC~349A & 1.7$^{22}$ && 60000$^{23}$ & 9$^{23}$ & $>70^\circ$$^{24}$ & 10$^{25}$ \\
S106 IRS & $0.5-1.7^{26}$ & O7.5V$^{26}$ && 13.6$^{27}$ & $\sim75^\circ$$^{28}$ & $\sim5^{29}$ \\
VV Ser & $0.2-0.6^{30}$ & B6V$^{2}$/B9V$^{31}$ &&$\sim3^{31}$ & $\gs80^\circ$$^{32}$ & $\sim10^{33}$\\
\end{tabular}

\label{Properties}
\caption{
  literature.  1: van den Ancker, de Winter \& Tjin A Djie (1998); 2: Hernandez
  et al.\ (2004); 3: Oudmaijer, Busfield \& Drew (1997), van den Ancker,
  Wesselius \& Tielens (2000); 4: Eisner et al.\ (2003, 2004); 5: Fuente et
  al.\ (1990) from NH$_3$; 6: Schreyer et al.\ (2002); 7: Willner et
  al.\ (1982); 8: Schreyer et al.\ (2006); 9: Waterlout \& Brand (1989) from
  molecular gas (CO); 10: Prisinzano et al.\ (2005) assuming it is at the same
  distance as the resolved stellar cluster found in M8; 11: Thronson,
  Loewenstein \& Stokes (1979); 12: Linz et al.\ (2009); 13: Bunn et
  al.\ (1995) from the star and Simon et al.\ (1984) from CO; 14: Estimates
  vary from 1.3kpc (Hanson, Howarth \& Conti 1997) using spectral parallaxes to
  2.1kpc (eg Hoffmeister et al.\ 2008) using photometric methods -- we adopt
  the latter; 15: Follert et al.\ (2010); 16: Porter et al.\ (1998); 17: Bunn
  et al.\ (1995) from the star, Lada (1976) from CO; 18: Drew et al.\ (1997);
  19: Alonso-Albi et al.\ (2009); 20: Acke et al.\ (2008); 21: Malbet et
  al.\ (2007); 22: Meyer, Nordsieck \& Hoffman (2002); 23: Cohen et
  al.\ (1985); 24: Danchi, Tuthill \& Monier (2001); 25: Gordon et al.\ (2001);
  26: Schneider et al.\ (2007), we adopt the further distance, which their
  analysis shows results in the given spectral type; 27: van den Ancker et
  al.\ (2000); 28: Hippelein \& Muench (1981), Solf \& Carsenty (1982); 29:
  Bunn et al.\ (1995); 30: Montesinos et al.\ (2009); 31: Pontoppidan et
  al.\ (2007); 32: Eisner et al.\ (2003); 33: Corcoran \& Ray (1997).  }

\end{table*}

\section{Observations}
The data were acquired on the nights of 24 and 25 June 2000 and 7 and 8 July
2001 using the facility near infrared spectrometer CGS4 on UKIRT.  We used the
echelle grating and 300mm camera with a one pixel wide slit.  Due to anamorphic
distortion within the instrument the spatial scale along and across the slit
vary.  The effective size of a pixel with this setup is 0.4 arcseconds across
the slit, and 0.91 arcseconds along it.  The measured spectral resolution from
an arc calibration line was 12kms$^{-1}$.  The slit was oriented at the default
north-south position angle.  We observed two separate wavelength ranges for all
of our targets, from approximately 1.636--1.648$\mu$m in order to detect the
1.6411725$\mu$m Br12 and 1.6440018$\mu$m a$^4$F$_9$--a$^4$D$_7$ [FeII] lines,
and 1.679--1.691$\mu$m  to detect the 1.6811164$\mu$m Br11 and
1.6877808$\mu$m c$^4$F$_9$--z$^4$F$_9$ FeII lines.
We also observed the 2.159--2.173$\mu$m range in order to measure
2.166127$\mu$m Br$\gamma$ for five of the targets.  All wavelengths are in
vacuum, and hereafter all wavelengths are truncated to three decimal places.
All eight of the main
targets are known massive YSOs or Herbig Ae/Be stars.  Some basic properties of
these are given in Table \ref{Properties}.  Where possible we have derived
inclinations from near or mid-infrared interferometry.  For GL~490 we instead
use results from mm CO interferometric measurements, and for S106 IRS we rely
on models of the velocity structure.  Where required the tabulated luminosities
have been modified from the original references to take account of revised
distances.  For sources where an accurate spectral type can be inferred we have
given that.  For those where only a bolometric luminosity exists (generally the
most heavily embedded) we have quoted that instead.  All of these sources were
chosen to be bright enough at $H$ to be observable with CGS4.  In addition, we
observed the planetary nebula NGC~7027 and the yellow hypergiant IRC~+10420 for
calibration purposes. NGC~7027 provided a check of our wavelength calibration,
since its velocity is known to a greater degree of accuracy than most of our
target stars.  IRC~+10420 was observed as it is a known FeII emitter in the
optical, and hence provides a check of our line identification.  The sample is
not in any sense complete or statistically representative of massive YSOs as a
whole.

\begin{figure*}
\centering
\includegraphics[width=6.5in,angle=0]{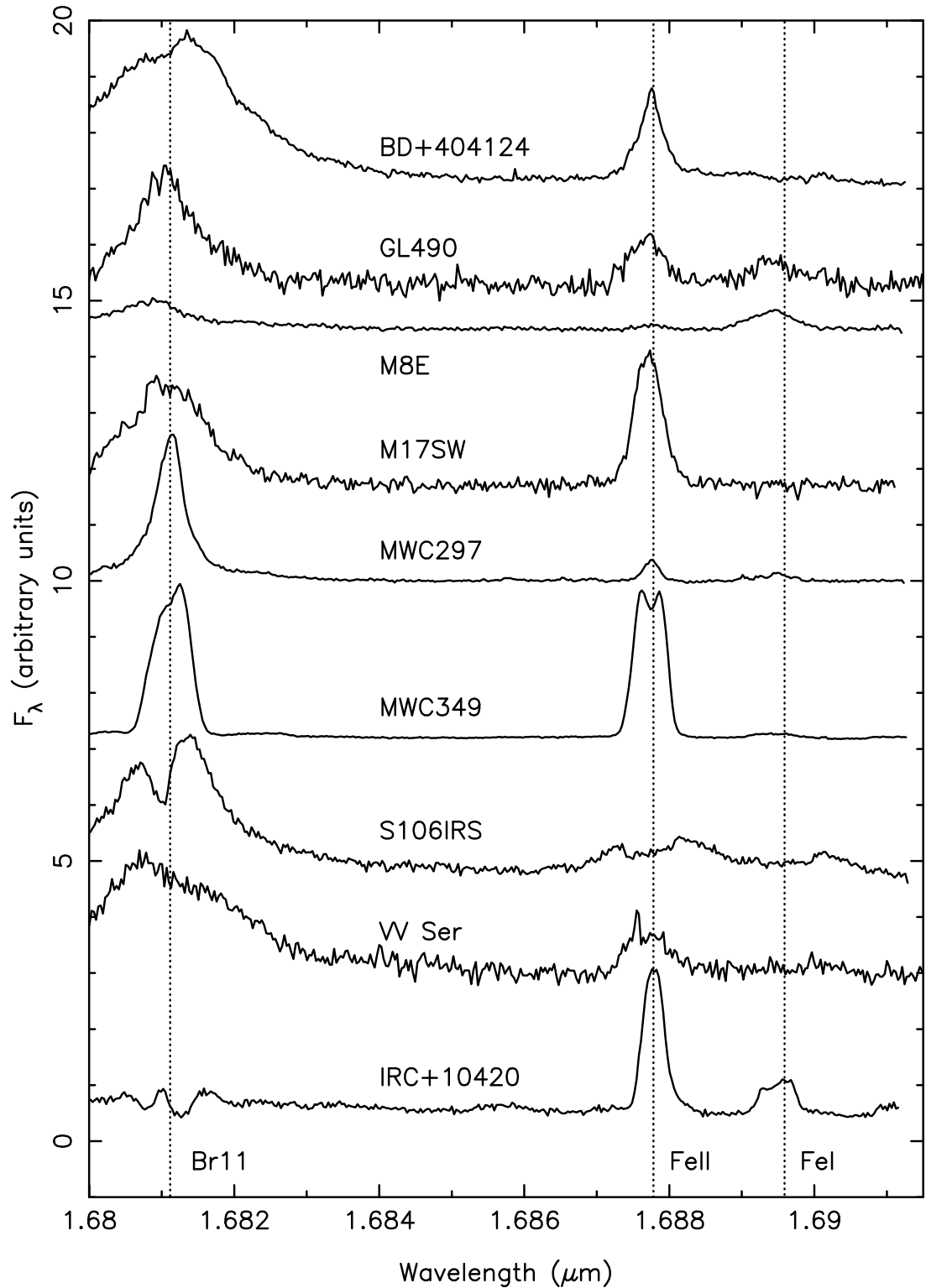}

\caption{ Spectra of the eight target stars and of the known FeII emitter
  IRC+10420 in the wavelength range showing HI Br11 1.681$\mu$m and the
  1.688$\mu$m z$^4$F$_9$--c$^4$F$_9$ FeII line.  Both the FeII line and HI
  Br11  rest wavelengths are indicated by the dotted lines, and the
  previously unidentified feature we associate with FeI is also shown.  The
  spectra are presented in the rest frame after correction for their
 published $v_{LSR}$ values for easier comparison.  } \label{fig1}
\end{figure*}

\begin{figure*}
\centering
\includegraphics[width=6.5in,angle=0]{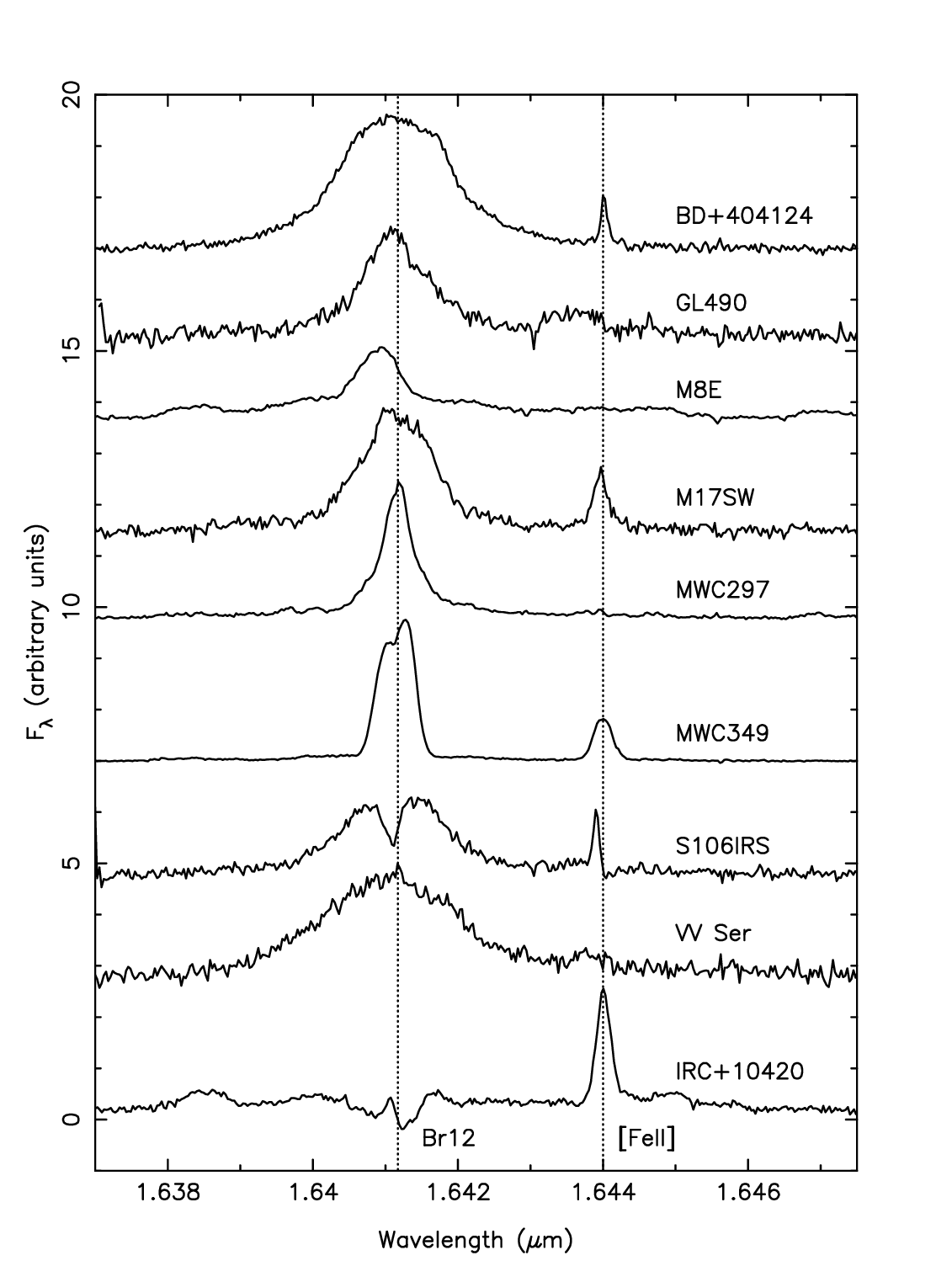}

\contcaption{ Spectra of the eight target stars and of the known FeII emitter
  IRC+10420 in the wavelength range showing HI Br12 1.641$\mu$m and 
  1.644$\mu$m a$^4$F$_9$--a$^4$D$_7$ [FeII].  Both the [FeII] line and HI
  Br12  rest wavelengths are indicated by the dotted lines.  The
  spectra are presented in the rest frame after correction for their
 published $v_{LSR}$ values for easier comparison.  }
\end{figure*}

All the sources are sufficiently compact that they could be observed by nodding
the telescope such that the sources remained on the slit (the measured beam
separation is 22 arcseconds). 
%
We observed standard stars to provide crude flux calibration and correct for
atmospheric absorption features in addition to observations of the targets.
The standard stars were chosen to have a range of spectral type from G6V
through to O9V.  This provided enough range to enable us to correct for the
observed Brackett line absorption in the standards themselves.  Cooler stars
were not used since they have many other spectral features in the wavelength
ranges observed, and in practice we used the hottest standard available for any
target.  The Brackett line absorption in the standard was interpolated over to
remove it.  This gives good though not perfect correction: Br11 in particular
may be affected by poor correction in the standard star since the line lies
near the edge of the spectrum (cf Figure \ref{fig1}).  There are some hints
that the Br11 and Br12 line profiles differ in some objects (eg
BD+40$^\circ$4124, Figure \ref{fig1}), but the difficulty in correcting
adequately for the instrumental response and interpolation of Br11 in the
standard star makes it impossible to be sure these are real.  We therefore do
not use Br11 in our analysis.  For Br12 and Br$\gamma$ the interpolation
process is far more secure, and is certainly adequate to define weak line
wings.

A standard reduction procedure was adopted for all of the data discussed here.
The projected image of the slit on the array is curved. This was corrected for
using the trace of a bright arc line as a model.  This procedure was checked
for reliability using fainter night sky lines present on each individual
exposure.  Each frame was also checked to ensure that no velocity shifts were
apparent, again using the night sky lines.  Where such shifts were seen the
data were realigned using the centroid of the night sky line.

Both of the beams were extracted from the data after correction for 
spectral distortion with wavelength.
The typical extraction is 6 spatial pixels (or
4.5 arcseconds) to encompass the full image of the spectrum as seen on the
array.  The telescope was slightly defocused to correct for problems caused by
the order sorting filter present in CGS4, which utilises a circular variable
filter (CVF).  This imposes a wavelength dependent transmission pattern on the
spectra that contains both a high frequency component (approximately one
resolution element) and a lower frequency component where the CVF response
declines away from the centre of the observed wavelength range, giving lower
transmission and signal-to-noise near the edges of the spectrum.  The high
frequency component also varies slightly with spatial position.  The best
correction is achieved if both object and atmospheric standard star have the
same point spread function (PSF), so that the pattern divides out during the
flux calibration.  The easiest way to ensure this with CGS4 was to smear the
PSF slightly for both.  The data from the separate beams were treated
individually since the CVF pattern is seen to vary at the few \% level between
the two beams.

The array was stepped six times across two pixels in order to fully sample the
resolution element, and reduce the impact of bad pixels.  These
individual frames must be interleaved to give the final data and it is possible
that any of sky transmission, atmospheric seeing or the position of the object
in the narrow slit can change between frames.  This is corrected for
by looking for appropriate periodicity in the final spectra (ie a pattern
repeating every 3 pixels) which can then be used to apply a correction
factor to the data.  Generally such corrections are small ($<$1\%) 
but important in obtaining the best possible signal-to-noise ratio from the 
acquired data.

Wavelength calibration was achieved using the telluric 
atmospheric absorption features
present in all spectra.  There are significantly more of these than either arc
lines or night sky lines and hence they provide a much better map of the
distortion of the wavelength scale and of the variation from object to object.
We measured the wavelengths of the absorption features from the much higher
spectral resolution atmospheric absorption spectrum published by Hinkle,
Wallace \& Livingston (1995).  Comparison of the different HI Brackett lines
showed this procedure gave consistently good results, allowing us to compare
the different lines with confidence.  Our observations of NGC~7027 indicated
our wavelength calibration should be good to within $\pm8$kms$^{-1}$.

Finally, after objects are ratioed with the standard stars the two beams are
combined to give the final flux calibrated spectrum.  The final spectra are
shown in Figure \ref{fig1}.

\section{Results}
\subsection{General Properties}

\begin{figure*}
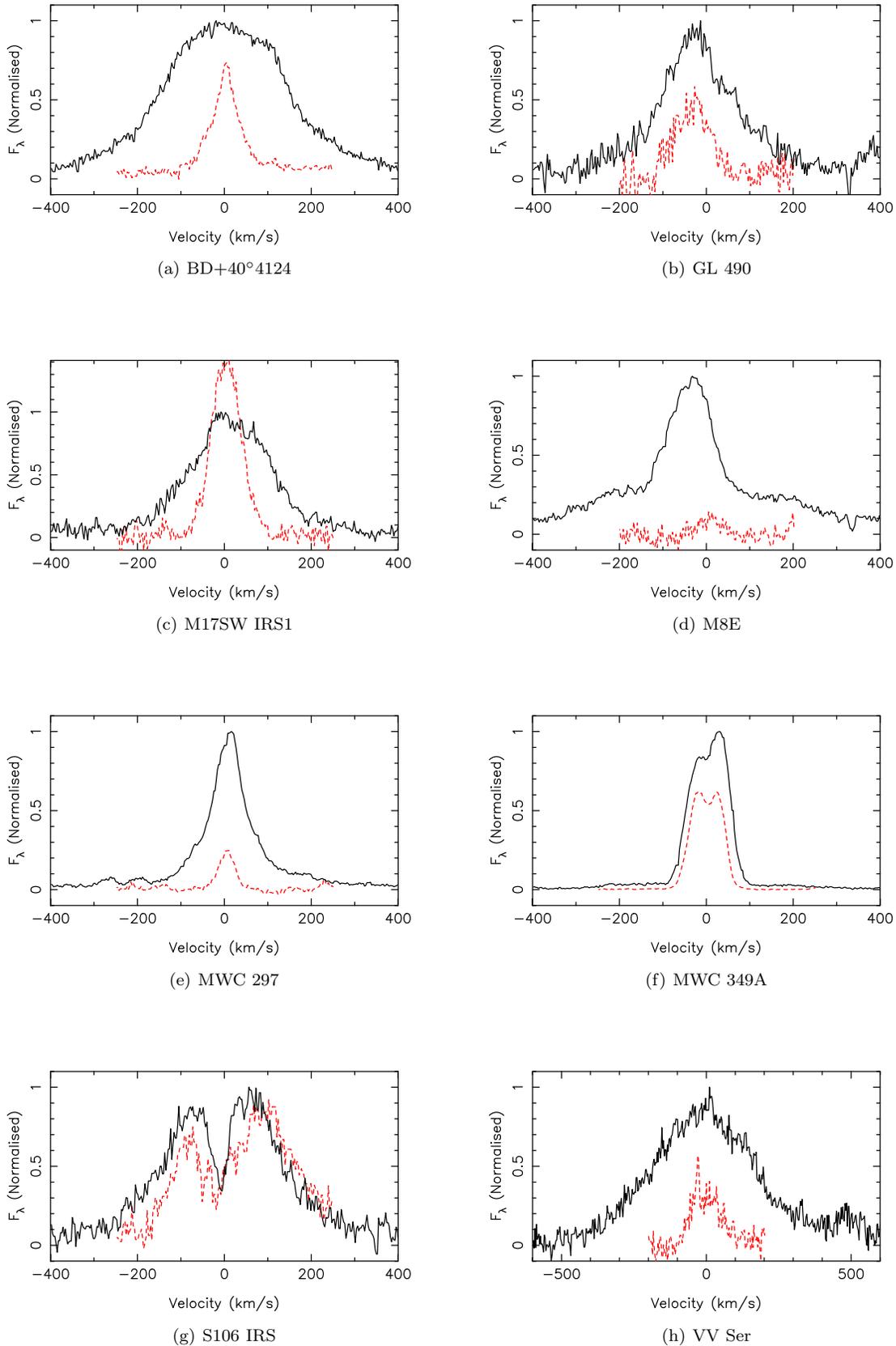

\centering
\subfigure[BD+40$^\circ$4124]{
\includegraphics[width=3in,angle=0]{fig2_1.ps}
\label{fig2a}}
\subfigure[GL~490]{
\includegraphics[width=3in,angle=0]{fig2_2.ps}
\label{fig2b}}

\subfigure[M17SW IRS1]{
\includegraphics[width=3in,angle=0]{fig2_3.ps}
\label{fig2c}}
\subfigure[M8E]{
\includegraphics[width=3in,angle=0]{fig2_4.ps}
\label{fig2d}}

\subfigure[MWC~297]{
\includegraphics[width=3in,angle=0]{fig2_5.ps}
\label{fig2e}}
\subfigure[MWC~349A]{
\includegraphics[width=3in,angle=0]{fig2_6.ps}
\label{fig2f}}

\subfigure[S106 IRS]{
\includegraphics[width=3in,angle=0]{fig2_7.ps}
\label{fig2g}}
\subfigure[VV Ser]{
\includegraphics[width=3in,angle=0]{fig2_8.ps}
\label{fig2h}}

\caption{ Br12 (solid) and FeII (dashed) line profiles after continuum
  subtraction.  The velocity scale for all objects is in their own rest frame,
  after correction according to the nominal $v_{LSR}$ values given in Table 1.
  The spectra are
  normalised so that the peak flux of Br12 is set equal to 1, and the FeII line
  is scaled relative to this.  } \label{fig2}
\end{figure*}

\begin{table*}
\tabcolsep=0.1Em
\begin{tabular}{lrcccrcccrccc}
Object & \multicolumn{4}{c}{Br12} &  \multicolumn{4}{c}{Br$\gamma$} &  \multicolumn{4}{c}{FeII} \\
 & \multicolumn{1}{c}{$v_{LSR}$}  & FWHM  & FWZI & Flux &
\multicolumn{1}{c}{$v_{LSR}$} & FWHM & FWZI&Flux   &
\multicolumn{1}{c}{$v_{LSR}$} & FWHM & FWZI  &Flux 
\\
BD+40$^\circ$4124 & $14\pm1$ & 320$\pm$30 & $\pm510$  & $1.61\pm0.05$
& $8\pm1$ &  $233\pm1$ & $-500/+530$  & $3.2\pm0.1$ 
& $3\pm1$ & $74\pm1$ &$\pm100$ & $0.29\pm0.01$ \\
GL~490 & $-24\pm1$ & $192\pm1$ & $\pm275$ & $0.4\pm0.02$ 
&& & & &
$-32\pm1$ & $97\pm1$ & $\pm80$ & $0.1\pm0.01$\\
M17SW IRS1 & $14\pm1$ & $210\pm20$ & $-420/+330$ & $0.3\pm0.02$ &
$8\pm1$ & $163\pm1$ & $\pm550$ & $1.3\pm0.01$ &
$4\pm1$ & $80\pm1$  &$\pm80$ & $0.1\pm0.01$\\
M8E & $-37\pm1$ & $116\pm3$ & $-600/+730$ & $0.9\pm0.05$ &
& && &
$9\pm1$ & $71\pm2$ & $\pm55$ & $0.02\pm0.01$ \\
MWC~297 & $10\pm1$ & $64\pm1$ & $-375/+400$ & $12.4\pm0.01$ &
& && &  
$6\pm1$ & $44\pm1$ & $\pm50$ & $1.06\pm0.03$ \\
MWC~349A & $7\pm1$ & $83\pm8$ & $\pm330$ & $24.0\pm0.1$ &
$15\pm1$ & $97\pm10$ & $-300/+400$ & $120\pm1$ &
$2\pm1$ & $68\pm7$ & $\pm100$ &$13.50\pm0.05$\\
S106 IRS & $-6\pm1$ & $280\pm30$ & $\pm420$ & $0.39\pm0.02$ 
& $-1\pm1$ & $185\pm20$ & $/+380$ & $4.54\pm0.01$ &
$19\pm1$ & $235\pm20$ & $-200/$ & $0.13\pm0.02$\\
VV Ser & $-18\pm1$ & $380\pm2$ & $\pm420$ & $0.52\pm0.1$ &
$-3\pm1$ & $320\pm5$ & $-600/+700$ & $2.1\pm0.1$ 
& $-6\pm1$ & $110\pm10$ & $\pm80$ & $0.07\pm0.01$ \\
\end{tabular}

\caption{Measured observed line parameters: all velocities are quoted in
  kms$^{-1}$, and all fluxes in units of 10$^{-15}$Wm$^{-2}\mu$m$^{-1}$.  The
  FWZI missing for the blue wing of Br$\gamma$ and the red wing of the FeII
  line in S106 IRS are due to strong residual line emission on the wings of the
  tabulated lines from other species.  The errors quoted are the nominal errors
  on the fit, and do not include any error in the velocity calibration itself.
}\label{SimpleFits}

\end{table*}

The hydrogen Brackett Br11 and Br12 lines were detected in
all objects observed, as is evident in Figure \ref{fig1} and Br$\gamma$ is also
present in all 5 sources we observed at that wavelength as shown in Figure
\ref{fig3}. We also detected the 1.688$\mu$m z$^4$F$_9$--c$^4$F$_9$ FeII in
all sources, though in M8E it is very weak.  Forbidden [FeII] emission was also
clearly present in BD+40$^\circ$4124, M17SW IRS1, MWC~349A and S106 IRS.  In
S106 IRS the [FeII] emission is extended, with a strong velocity gradient along
the full length of the slit. The bulk of the emission for our slit position
angle comes from the north-eastern cone of the bipolar nebula (Solf \& Carsenty
1982), and is blueshifted with regard to the systemic velocity. For each of
these four objects the [FeII] line width was 50kms$^{-1}$ or less.  The [FeII]
line is a known tracer of fast shocks, such as seen in collimated outflows near
young stellar objects, and the line widths observed are perfectly consistent
with this explanation.  There is also a weak broad feature in both GL~490, and
possibly VV Ser, that may be due to [FeII] but is offset to the blue from the
systemic velocity (by about 60kms$^{-1}$ and 10kms$^{-1}$ respectively).
However the width of the line in these cases, about 100kms$^{-1}$ for VV Ser
and 200kms$^{-1}$ for GL~490, suggests a different origin from the other
sources unless we are seeing a one-sided jet coming towards us in these
sources.  These are the also two lowest luminosity sources we observed.  We
have been unable to identify any possible alternative identification, but it
seems probable that the line seen is not [FeII].  With these two exceptions,
the forbidden [FeII] emission has a very different line profile from the FeII
line, and is always much narrower than any of the other lines observed.  In
addition, the [FeII] line is seen in NGC7027 whilst the fluorescent FeII is
not, as expected for a purely nebular origin for the emission in this source.
We can be confident therefore that the [FeII] line has an origin in the
extended nebular gas around our sources when present, and FeII line has a
different origin.  We will not consider the [FeII] line further in this paper
since we primarily observed it to demonstrate that the fluorescent FeII
emission was not excited in a similar fashion to the [FeII] emission.

An otherwise previously unidentified line at $\sim$1.68955$\mu$m is clearly
visible in GL 490, M8E, MWC~297 and MWC~349A, and possibly in S106~IRS.  In
S106~IRS the width and double peaked nature of the main FeII line blends with
this feature, making the exact identification difficult, but there is excess
red emission present that would naturally be explained if this unidentified
line was also present here with a similar profile to that of Br12 and FeII.
Both FeI and CI lines are identified at 1.68959$\mu$m in the solar spectrum
(Ramsauer, Solanki \& Biemont 1995).  Observations of FeI 8387\AA\ in IRC+10420
(eg.\ Oudmaijer 1998) show an almost identical line profile to the line shown
here in that object (Figure \ref{fig1}).  We therefore prefer an identification
with FeI for this line.  Another unidentified line can be seen near
1.6385$\mu$m in IRC+10420 and possibly M8E, which again might be identified
with a blend of two FeI lines (from the atomic data presented in Mel\'{e}ndez
and Barbuy 1999).  This is certainly consistent with its appearance in only
these two objects, but the profile is of poorer signal-to-noise in IRC+10420,
which, together with the possible blending, makes it difficult to compare the
profile of this line with that seen at 1.68959$\mu$m.

Figure \ref{fig2} and Figure \ref{fig3} present the key observational results
that we will interpret in the rest of this paper, namely a comparison of the
profiles of the FeII, Br12 and Br$\gamma$ lines.  The continuum has been
subtracted from all the profiles.  We note that weak HeI line emission is
present in both S106~IRS and MWC~349A, on the blue wing of Br12 and, more
obviously, Br$\gamma$, but we have not attempted to correct for this.  We have
corrected the observed lines for extinction in order to make a detailed
comparison.  We adopt the extinction law from Cardelli, Clayton \& Mathis
(1989) to convert the literature values in Table \ref{Properties} to $A_H$ and
$A_K$, using the standard value of $R=3.1$.  The data shown in Figure
\ref{fig2} and Figure \ref{fig3} have had these corrections applied.  Finally
we have normalised the plotted data such that the peak intensity of Br12 is
given a value of unity, and all other corrected fluxes scaled relative to this.

In order to compare the data with previously published work, and as a crude
guide to the properties reflected by the observed line emission, we have also
fitted all of the lines with a Gaussian.  This was done to determine the
systemic velocity as measured by that line, $v_{LSR}$, and the approximate full
width at half maximum (FWHM).
Where double peaked line profiles are present single Gaussian fits were used to
measure line centre and FWHM, with the fit restricted to the line wings.  This
gives reasonably accurate values for $v_{LSR}$ but the line width is only
accurate to approximately 10\%, and the quoted error reflects this.  The 
full width at zero intensity (FWZI)
is measured from the point where the line is 3$\sigma$ above the noise level,
after smoothing the data by a factor of two, for direct comparison with the
data presented in Bunn et al.\ (1995).  The results are shown in Table
\ref{SimpleFits}.

Finally we note that the objects are split almost evenly between those where
the overall general line shape is similar in Br12 and FeII, and those which show
significant differences.  However, 
the hydrogen lines are as broad or broader than the FeII line as measured at the
half maxima points, and in BD+40$^\circ$4124, GL~490, M17SW IRS1 and VV Ser the
difference is a factor of 2--3 as is easily seen in Figure \ref{fig2} and Table
\ref{SimpleFits}.  The
hydrogen lines also have much higher velocities in the line wings as measured
at zero intensity, with the possible exception of S106 IRS.  In all cases the
velocities observed are too large to be explained by gas excited in a standard
HII region.  There is a rough trend for more inclined objects to have broader
Br12 lines, as would be expected when the motions in the disk plane 
are more along our line of sight (eg Horne and Marsh 1986).

%
%
%

\begin{figure*}
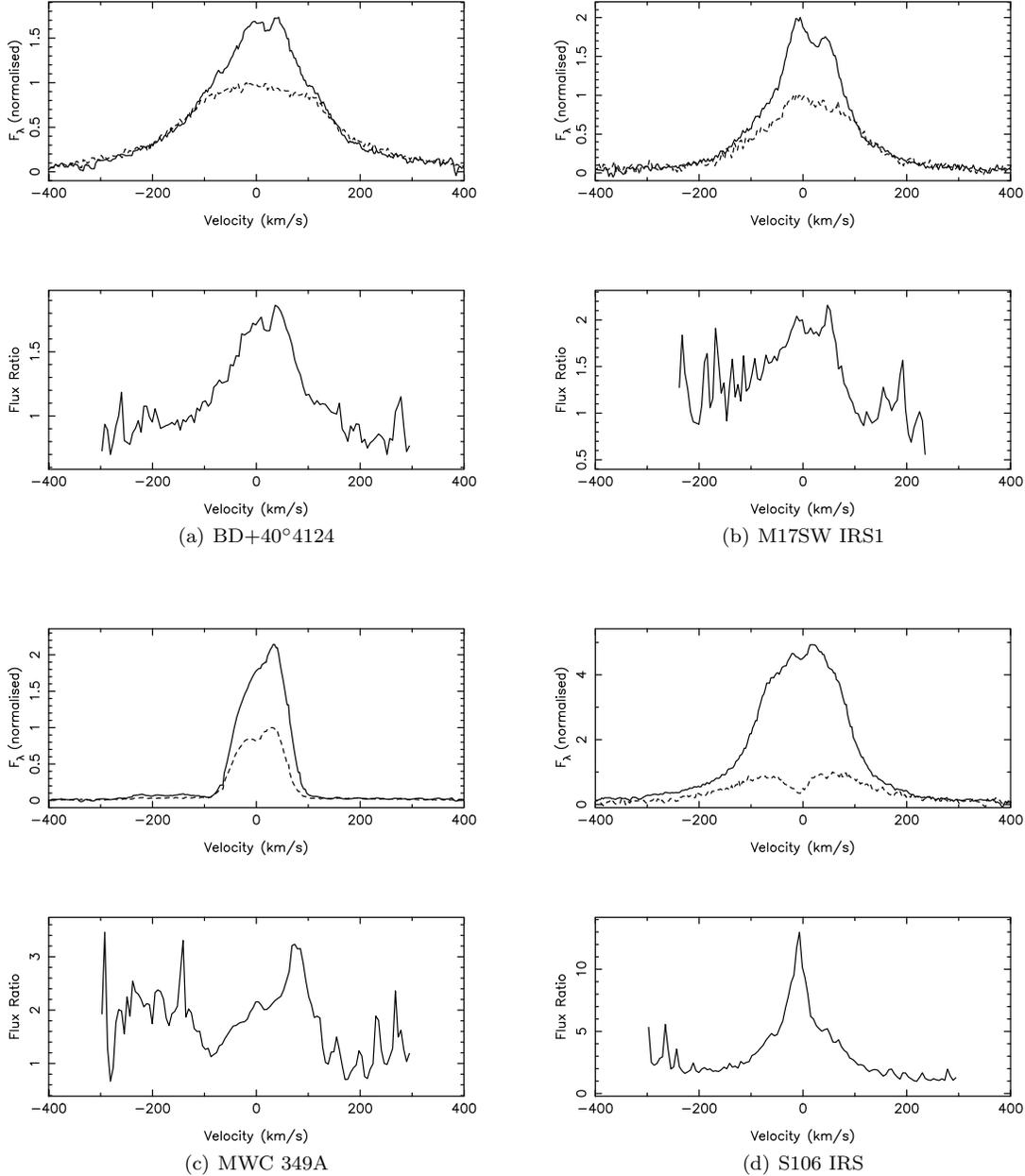

\centering
\subfigure[BD+40$^\circ$4124]{
\includegraphics[width=3.2in,angle=-90]{fig3_1.ps}
\label{fig3a}}
\subfigure[M17SW IRS1]{
\includegraphics[width=3.2in,angle=-90]{fig3_2.ps}
\label{fig3b}}

\subfigure[MWC~349A]{
\includegraphics[width=3.2in,angle=-90]{fig3_3.ps}
\label{fig3c}}
\subfigure[S106 IRS]{
\includegraphics[width=3.2in,angle=-90]{fig3_4.ps}
\label{fig3d}}

\caption{ Br$\gamma$ and Br12 line profiles after continuum subtraction, and
  correction for extinction.  The Br12 line has been scaled as in Figure
  \ref{fig2}, and the Br$\gamma$ lined scaled relative to this.  HeI
  7$^{3,1}$G--4$^{3,1}$F emission is evident in MWC~349A between a velocity of
  $\sim-250\rightarrow-100$kms$^{-1}$ on the Br$\gamma$ spectrum.  The same
  emission is also present in S106~IRS though only shows because of the
  asymmetry of the blue and red line wings to Br$\gamma$. We have not attempted
  to correct the plotted ratio for this HeI emission.  The ratio shown is 
  for the line profiles after the corrections stated above have been applied.
  The ratio is smoothed by a factor
  of two relative to the line profiles, and truncated at the point
  where noise becomes dominant as noted in the text.  The velocity scale is in
  the rest frame of the object as in Figure \ref{fig2}.  } \label{fig3}
\end{figure*}

\begin{figure*}
\centering
\includegraphics[width=3.3in,angle=-90]{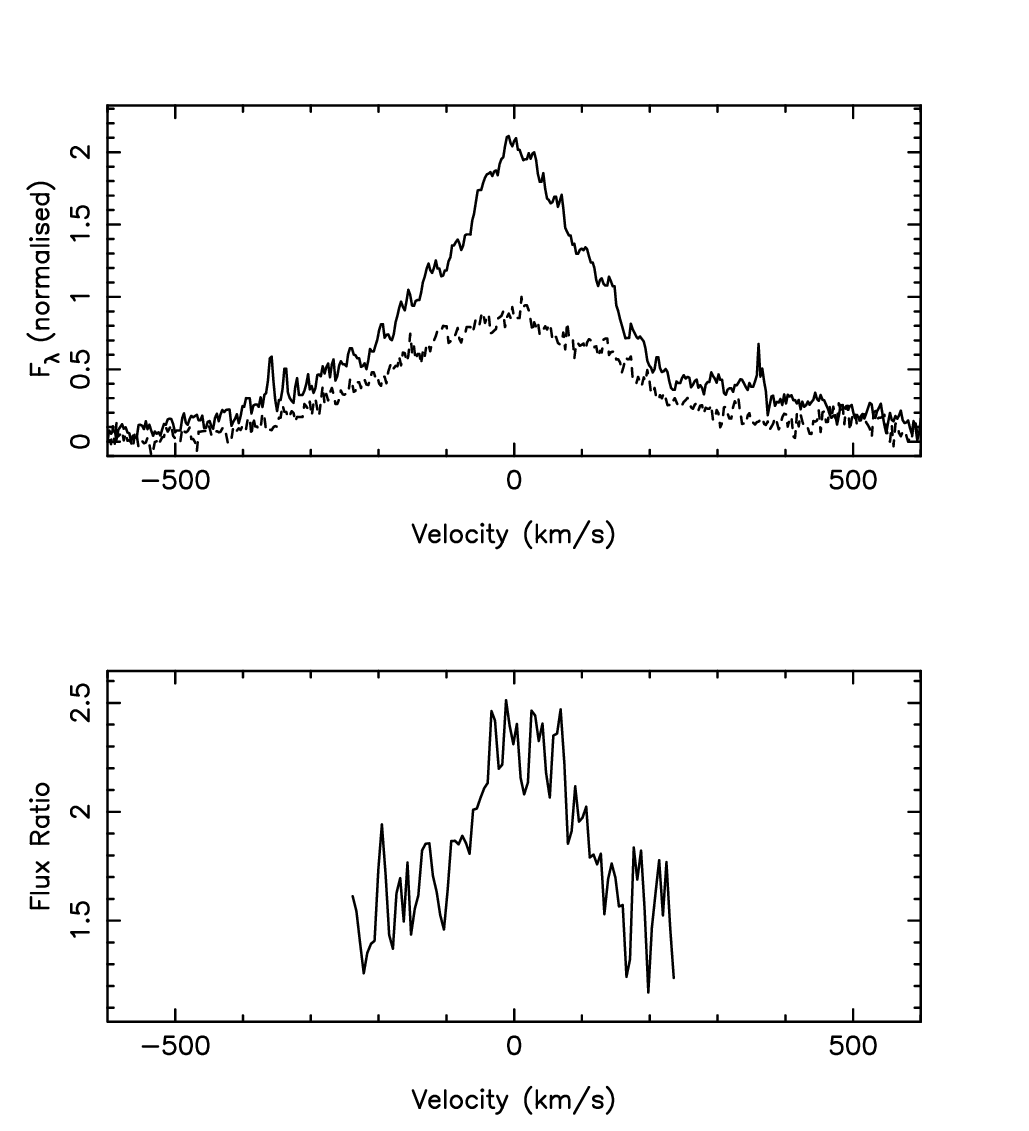}

\contcaption{ (e) VV Ser
} \label{fig3e}
\end{figure*}

\subsection{Comparison of the hydrogen line profiles}
\label{hydrogen}
The comparison of velocity resolved hydrogen line ratios can potentially reveal
the origin of the emission.  If the gas is optically thin, this ratio will be a
constant and will tend towards the case B ratio (eg Storey \& Hummer 1995).
Bunn et al.\ (1995) however showed it was possible to define the origin of the
emission in the case where the lines are at least partially optically thick.
Br11 and Br12 are too close together to provide a useful comparison of line
opacity (and as noted above we discount the Br11 lines from analysis given the
problems associated with its location near the edge of the observed spectra).
We did observe BD+40$^\circ$4124, M17SW IRS1, MWC~349A, S106 IRS and VV Ser in
both Br$\gamma$ and Br12 lines.  Figure \ref{fig3} shows the Br$\gamma$ and
Br12 lines for each of these objects, corrected for the inferred extinction as
described in Section 3.1, as well as the ratio of these lines.  Away from the
bright line cores, the main source of noise is actually non-Gaussian read-noise
from the array.  We have therefore truncated the line ratio plots at the
velocity where the noise starts to dominate the ratio, but it is evident from
the over-plot of the lines themselves that the ratio remains roughly constant
even beyond this.  The presence of HeI emission is evident in MWC~349 and
S106~IRS, at a systemic velocity of $\sim-200$kms$^{-1}$ relative to
Br$\gamma$, as a turn-up in the ratio at that point. 

The ratio in absorption cross-section between Br$\gamma$ and Br12 is 11.1,
whilst that between Br$\alpha$ and Br$\gamma$ is 15.9, and Br$\alpha$ and
Pf$\gamma$ is 14.0.  The comparison of Br$\gamma$ and Br12 is therefore a
valuable test of the line formation mechanism in the same sense as the
comparison of Br$\alpha$ and Pf$\gamma$ or Br$\gamma$ in the work of Bunn et
al.\ (1995).  For the two objects we have in common with Drew et al.\ (1993)
and Bunn et al.\ (1995), the line ratios actually show very similar results
near the systemic velocity (Figures \ref{fig3b} and \ref{fig3d}).  The earlier
data have slightly lower spectral resolution and are also of lower
signal-to-noise, which together can explain any differences seen.

The optically thin case B value of the Br$\gamma$/Br12 ratio is $\sim5\pm1$ for
values of the electron density in the range $10^4-10^9$cm$^{-3}$ and
temperature in the range $3000-30000$K (Storey \& Hummer 1995).  Simple theory
indicates that for completely optically thick lines the ratio of the
intensities of the two should scale simply as $I_1/I_2 =
(\lambda_2/\lambda_1)^4 (S_1/S_2)$ in the Rayleigh-Jeans limit, where $S_1$ is
the effective surface area of the emitting region of line 1.  {\em If} the
emitting region of the two lines has the same surface area then the limiting
ratio is $\sim$0.36 for Br$\gamma$/Br12.  In practice we would expect the lower
excitation Br$\gamma$ line to emit from a larger volume, and hence larger
surface area, so a value somewhat above this would be expected for gas which is
close to optically thick.  Figure \ref{fig3}  shows all five objects,
with the exception of gas near the systemic velocity in S106~IRS, lies between
the case B and optically thick limit, indicating opacity in at least one of the
Brackett lines.  Overall the trend in the ratios is very similar for four of
the objects, namely a largely symmetrical profile that peaks near systemic.
The exception is MWC~349, which has a very asymmetrical profile.  We therefore
defer a full discussion of a suitable model for MWC~349 to Section
\ref{m349comments}.

The generic features that any theory for the origin of the hydrogen line
emission must be able to explain can be summarised as follows.  First, we
reiterate that all of our sources show all of the observed hydrogen lines in
emission, and the objects show evidence for nearly symmetric very broad line
wings (Figures \ref{fig2} and \ref{fig3}).  Secondly, many of the line ratios
tend to be closer to the optically thick limit ($F_{Br\gamma}/F_{Br12}\sim1$)
in the line wings than in the core.  Last, three of the objects show evidence
for weak double peaked or self-absorbed line profiles, in either Br12 or
Br$\gamma$ or both.  In two of the sources the minimum is significantly offset
from the systemic velocity: in BD+40$^\circ$4124 it lies at $\sim21$kms$^{-1}$,
and in M17SW IRS1 at $\sim25$kms$^{-1}$.  In S106~IRS however it lies at
$\sim-3$kms$^{-1}$.  This however is within the error on our velocity scale,
and hence is also in practice consistent with a small redshift.  The first two
are suggestive of self-absorption due to infalling material.  The latter may
still be self-absorption, or an intrinsic double peaked profile centred on the
systemic velocity.  These three objects are either at intermediate
inclination angles or close to edge-on (Table \ref{Properties}).

These characteristics are shared by at least some lower mass pre-main sequence
stars as well.  Folha \& Emerson (2001) present Br$\gamma$ spectra of a large
number of T Tauri stars.  Symmetric broad wings and weak Br$\gamma$
line self-absorption are common in the objects they class as type I under the
scheme of Reipurth, Pedrosa \& Lago (1996: type I objects in the original work
refer to objects with symmetrical and unabsorbed H$\alpha$ profiles).  These
are all objects that show greater evidence of veiling in the continuum, which
is usually taken to indicate a high accretion rate.  Objects without
this veiling show less evidence for strong broad emission lines.  It
is worth noting that many of the strong T Tauri Br$\gamma$ line emitters still
show P-Cygni style wind absorption, but only in the HeI lines that end in the
metastable $2^{1,3}S$ levels (eg for the 1.083$\mu$m $2^3S-2^3P$ line, Edwards
et al 2006). Many Herbig Ae/Be stars also show pronounced P-Cygni profiles for
the HeI 1.083$\mu$m line (Oudmaijer et al.\ 2011) as does MWC~297 (Drew et
al.\ 1997), and another object in our sample, S106~IRS, has a P-Cygni profile
in the 2.058$\mu$m HeI $2^1S-2^1P$ line (Drew et al.\ 1993).  

The natural explanation for the HeI profiles is an outflowing wind.  The
terminal wind velocity is then given in the usual fashion by the blueshifted
absorption.  For both of the objects in the sample under consideration here,
S106~IRS (Drew et al.\ 1993) and MWC~297 (Drew et al.\ 1997), this gives a
probable terminal velocity of $\sim200$kms$^{-1}$.  The wings on the hydrogen
lines are still broader than this however (cf Figure \ref{fig3}).  In addition
the wings have approximately the same full width at zero intensity in Br12 as
Br$\gamma$.  As noted by Bunn et al.\ (1995), if an {\em accelerating} wind is
present the expectation is that the Br$\gamma$/Br12 line ratio should increase
at high velocities: the faster gas lies physically farther out, in less dense
gas, and is therefore optically thinner, and therefore the full width at zero
intensity should reduce for higher series members.  None of our observed ratios
match this.  We can therefore rule out a standard hot star accelerating wind as
the source of the emission.   A  decelerating (perhaps mass loaded)
or constant velocity wind cannot be ruled out however.

The obvious alternative would be a variant on a disc model.  However, this
would be expected to show double peaked line emission if edge-on and relatively
narrow emission without broad wings if face-on for a pure disc model.  A
combination of wind and disc might appear to solve the problem, but such a
geometry effectively masks much of the receding emission in stars that are near
face-on and the symmetry of the lines argues against this being the sole
mechanism responsible (see Section \ref{m297comments}).  An interesting
theoretical study with regard to the modelling of T Tauri stars, by Kurosawa,
Romanova \& Harries (2011), makes many of these same points about the
difficulty of fitting profiles to the strong HeI absorption, Br$\gamma$
emission sources.  Their model includes potential emission from a stellar wind,
a disc wind and magnetospheric accretion.  They are able to reproduce the need
for blueshifted absorption in HeI, and mostly symmetric emission in the
infrared hydrogen lines.  However, the lines are too narrow, and at high mass
accretion rate the hydrogen lines start to show inverse P-Cygni absorption from
the infalling gas, suggesting some extra component is required to fully explain
the broad wings in the hydrogen lines.  Notably, the disc wind also must have a
large opening angle to avoid significant self absorption even at relatively low
accretion rates, and the same is likely to be true for our sources.

This raises the possibility that additional line broadening may be present in
the hydrogen lines.  Typical densities of the inner regions of an outflowing
wind in these sources are naively of order $\gs10^{12}$cm$^{-3}$ assuming the
mass loss rate is $\gs10^{-6}$\Msolar/yr, the observed wind velocity of
$\sim200$kms$^{-1}$, and typical main sequence OB star radii.  The radius may
actually be an underestimate since models predict the stars are swollen until
they reach a final main sequence configuration (eg Hosokawa \& Omukai 2009),
consistent with the low stellar wind speeds we see.  However, the mass loss
rates are also likely to be substantially higher, so the density estimate is
relatively unchanged.  This density is large enough that Stark broadening can
become significant for some lines (cf Muzerolle et al 2001 for the case of T
Tauri stars).  Repolust et al.\ (2005) modelled the Stark broadened Brackett
line profiles for main sequence OB stars.  Although the winds present in main
sequence stars are very different, the density is similar to the estimate given
above for our simple outflowing wind, and the principle of whether broadening
occurs under such circumstances is identical.  Figure 3 in Repolust et al.\
shows that in this density range wings are easily generated by Stark broadening
for Br10, with broad wings similar to that seen in our data.  Electron
scattering can also be apparent with such high densities.  With
$T_e\sim10000$K, line wings comparable to the thermal velocity width of the
electrons of 550kms$^{-1}$ is possible.  We know scattering is significant
in some Herbig Be stars, including MWC~297 (eg Oudmaijer \& Drew 1999).
Another key advantage of a broadening mechanism is that it can help to fill in
any depression in the line caused by line self-absorption due to shadowing from
a disk (eg see the schematic in Malbet et al.\ 2007 with regard to MWC~297).
Finally, broadening decouples the line ratio at high velocity from the need for
an optically thick emission region at that velocity.

In principle we can even tell which broadening mechanism is actually present.
A simple comparison of the ratio of the intensity of the line wing to the peak
intensity will show stronger wings for higher series transition if Stark
broadening is present, since these are the states most affected by
perturbations due to neighbouring ions.  We would expect the opposite to be
true for electron scattering, where essentially it is the population of the
excited state that determines the likelihood of scattering.  Thus a line like
H$\alpha$ is more likely to have wings enhanced by scattering, and the
relatively high Br series lines we are studying are more susceptible to Stark
broadening.  Stark broadening should also tend, in the high velocity limit, to
wings that scale as $v^{-5/2}$ (Repolust et al.\ 2005).  The core of the line,
as shown clearly in the data of Repolust et al., should also be almost flat in
a $\log F_\lambda$ versus $\log v$ plot.  Scattering has been modelled
previously in the young stellar object Lk~H$\alpha$~101 by Hamann and Persson
(1989), using the prescription of Castor, Smith and van Blerkom (1970).  In
this model the scattering all occurs in a slab that lies just above the line
source (the stellar wind), and is optically thin to Thomson scattering.  Laor
(2006) describes a model for scattering in the broad line region in active
galaxies, with a similar requirement on the opacity but an isotropic
illumination of the electrons, rather than a one sided illumination.  The
resultant functional forms for the scattered flux are different, but both give
profiles that are roughly exponential for likely values of the two key
scattering parameters (opacity and electron velocity).  In principle, the
Castor et al.\ model can be used to measure these scattering parameters, and
hence also quantities such as density at the base of the wind.  In practice,
this has not proved possible as we shall discuss in our analysis of the best
fit to the line profiles in Section 3.4, and in Section 4.

Finally, as noted by Drew et al.\ (1993), an optically thin nebular component
may be required in some cases in addition, of which S106~IRS is a clear example
given it powers an extended bipolar HII region (all three Brackett lines
presented are visible across an extended region on the long slit spectrum, with
lines of width $\sim30$kms$^{-1}$).  In that instance the line ratio at
systemic velocity is actually above the case B value.  This most likely
indicates that the optically thin nebular gas actually has lower extinction
than S106~IRS itself (ie we have over-corrected for the extinction in this
component), which is consistent with the fact the nebula is visible in the
optical and the star itself is extremely faint.

Overall therefore, broadening gives a natural explanation for the wings in all
our sources.  If we accept that factor, then a decelerating, constant wind or
large angle disc wind are viable explanations for the core of the lines.  This
is probably a natural explanation for all massive and intermediate
mass young stellar objects since none of our objects are selected to be in any
way ``special''.

\begin{figure*}
\centering
\subfigure[BD+40$^\circ$4124]{
\includegraphics[width=2.7in,angle=0]{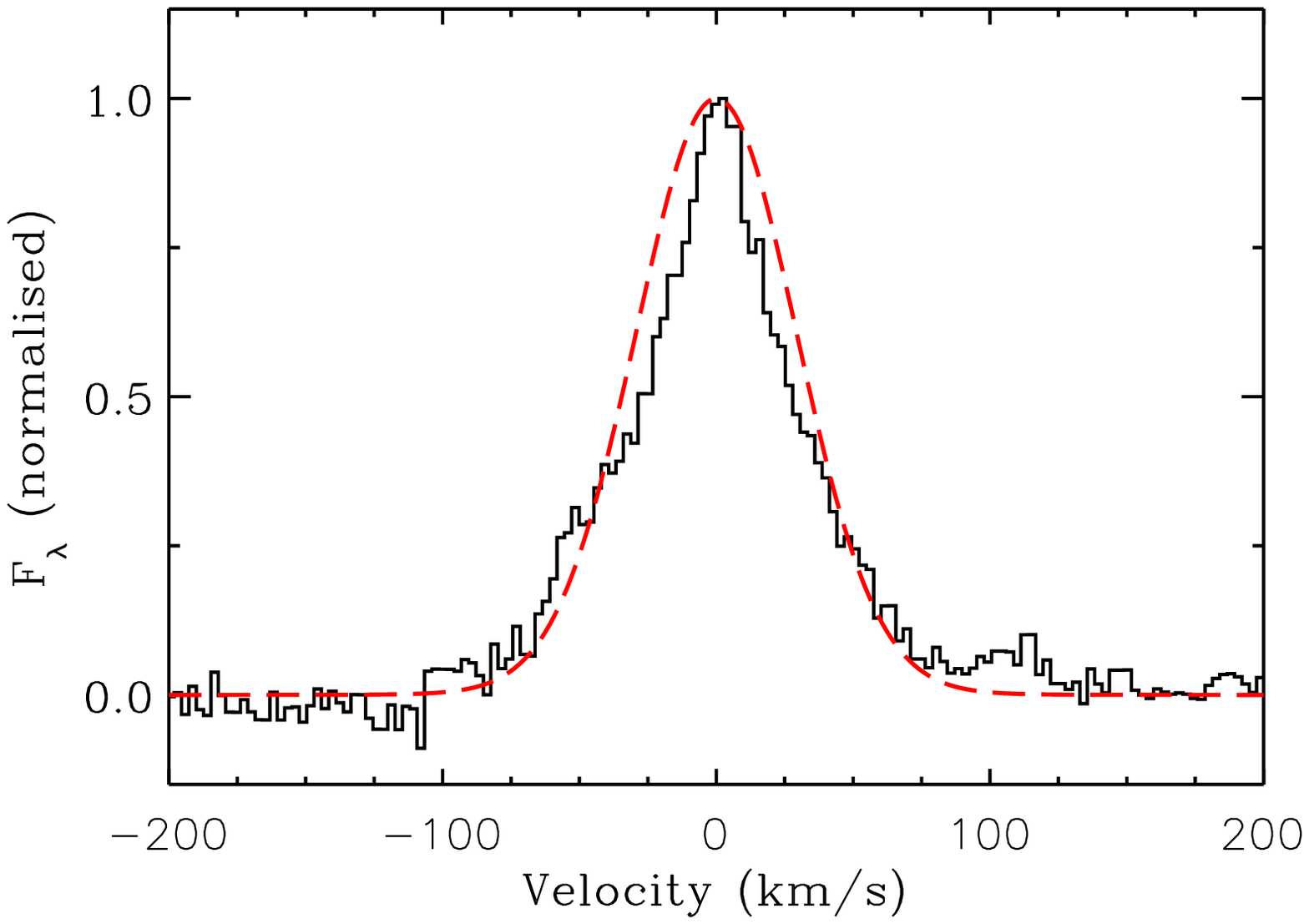}
\label{fig4a}}
\subfigure[GL~490]{
\includegraphics[width=2.7in,angle=0]{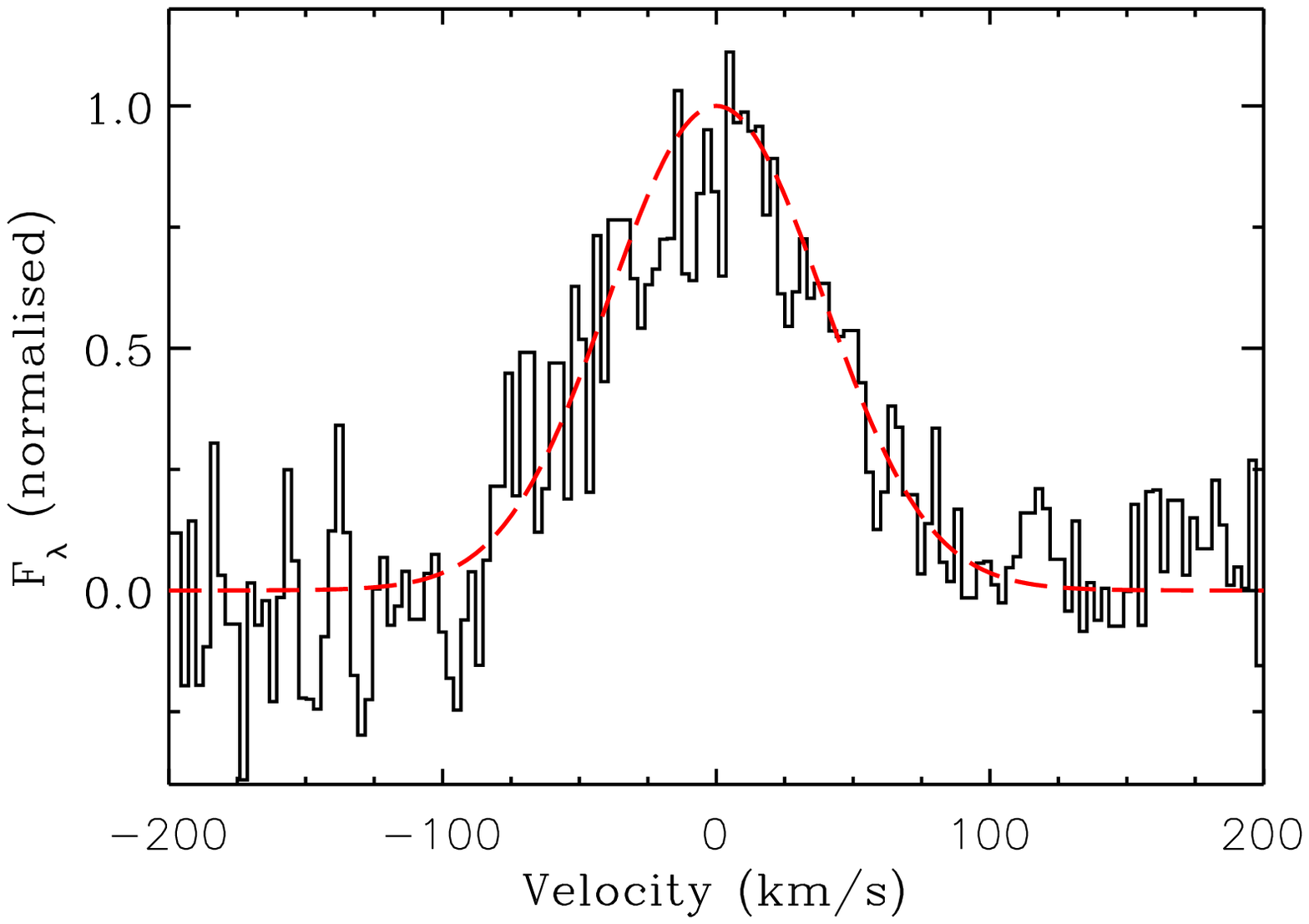}
\label{fig4b}}

\subfigure[M17SW IRS1]{
\includegraphics[width=2.7in,angle=0]{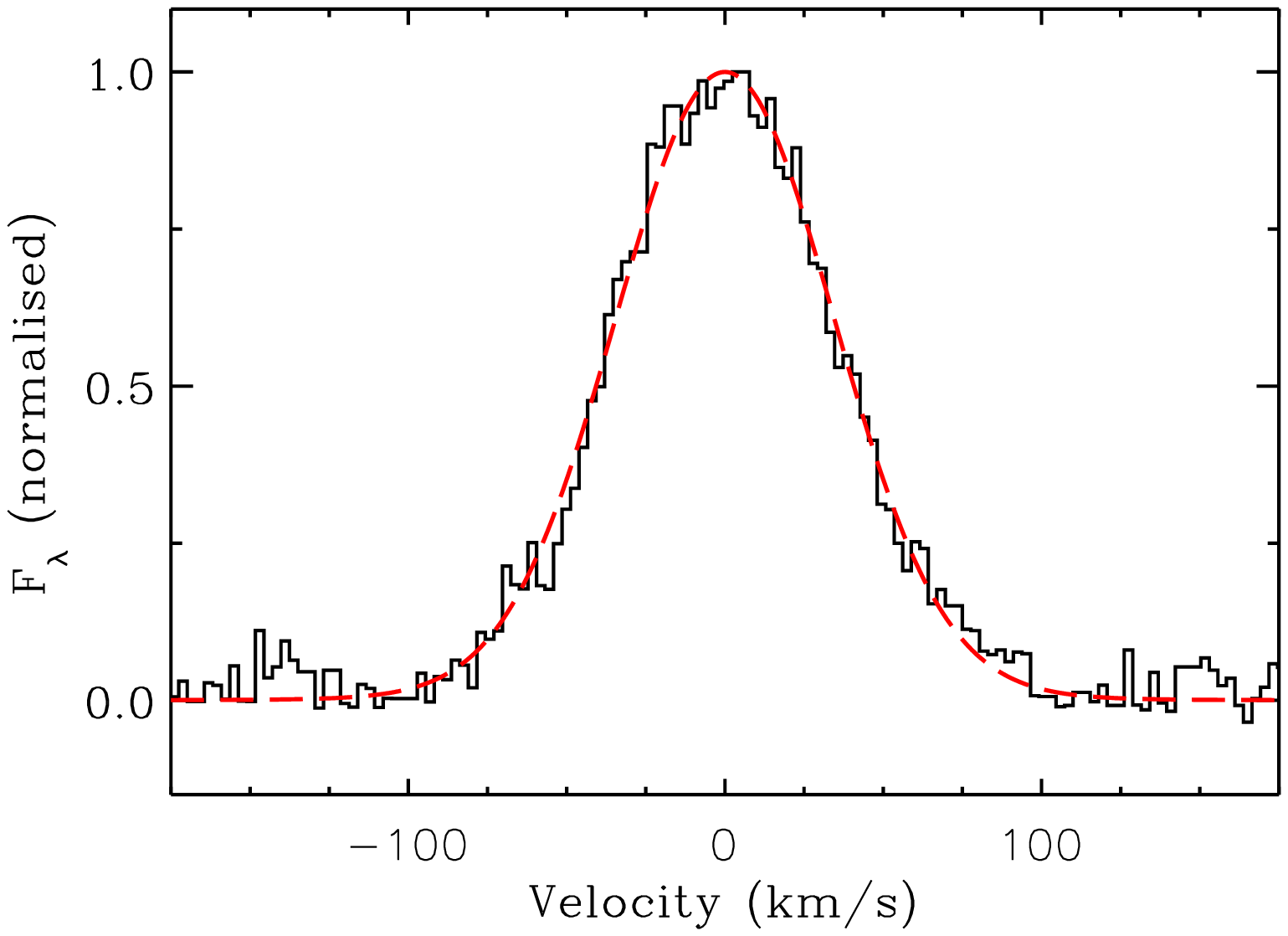}
\label{fig4c}}
\subfigure[M8E]{
\includegraphics[width=2.7in,angle=0]{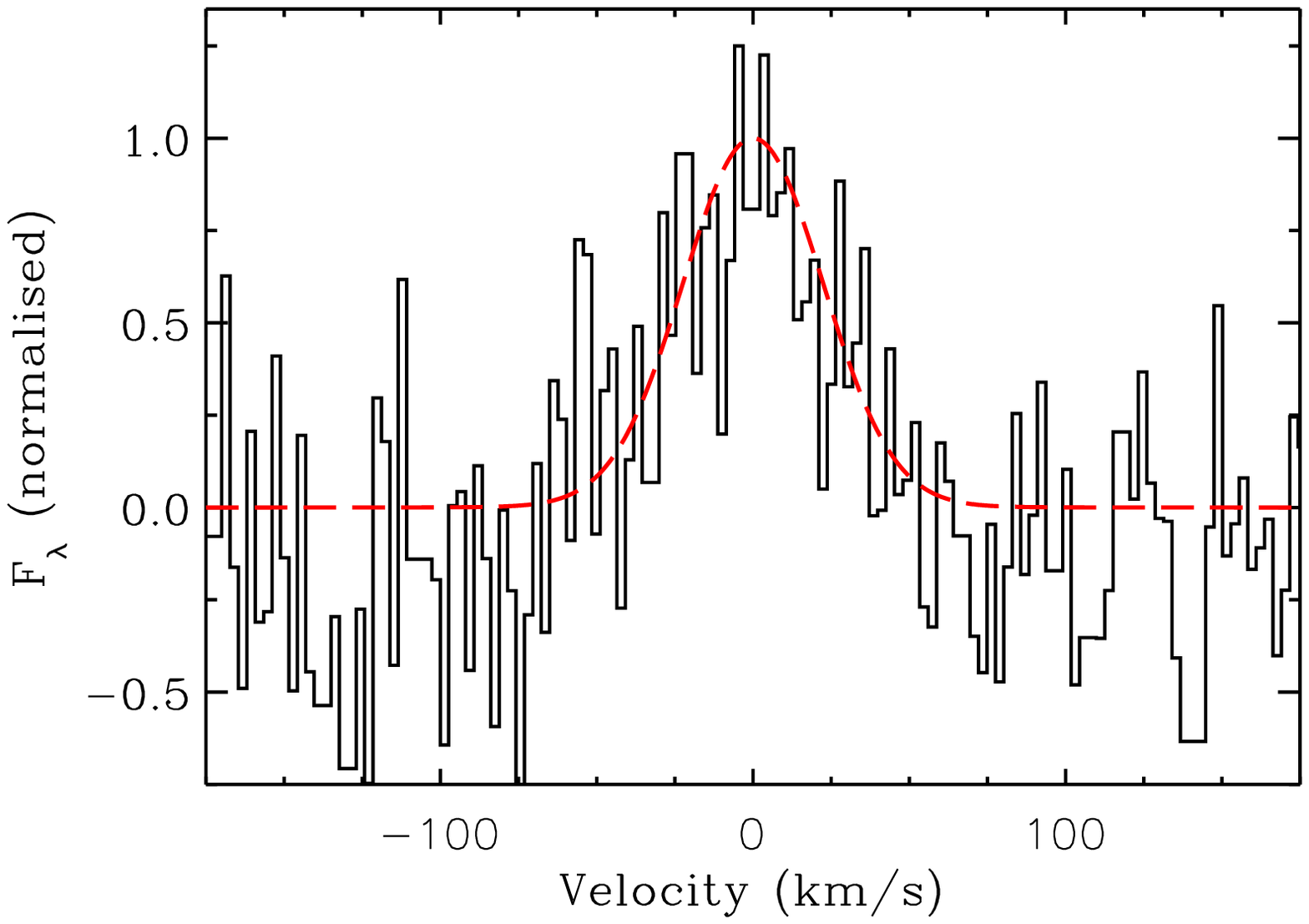}
\label{fig4d}}

\subfigure[MWC~297]{
\includegraphics[width=2.7in,angle=0]{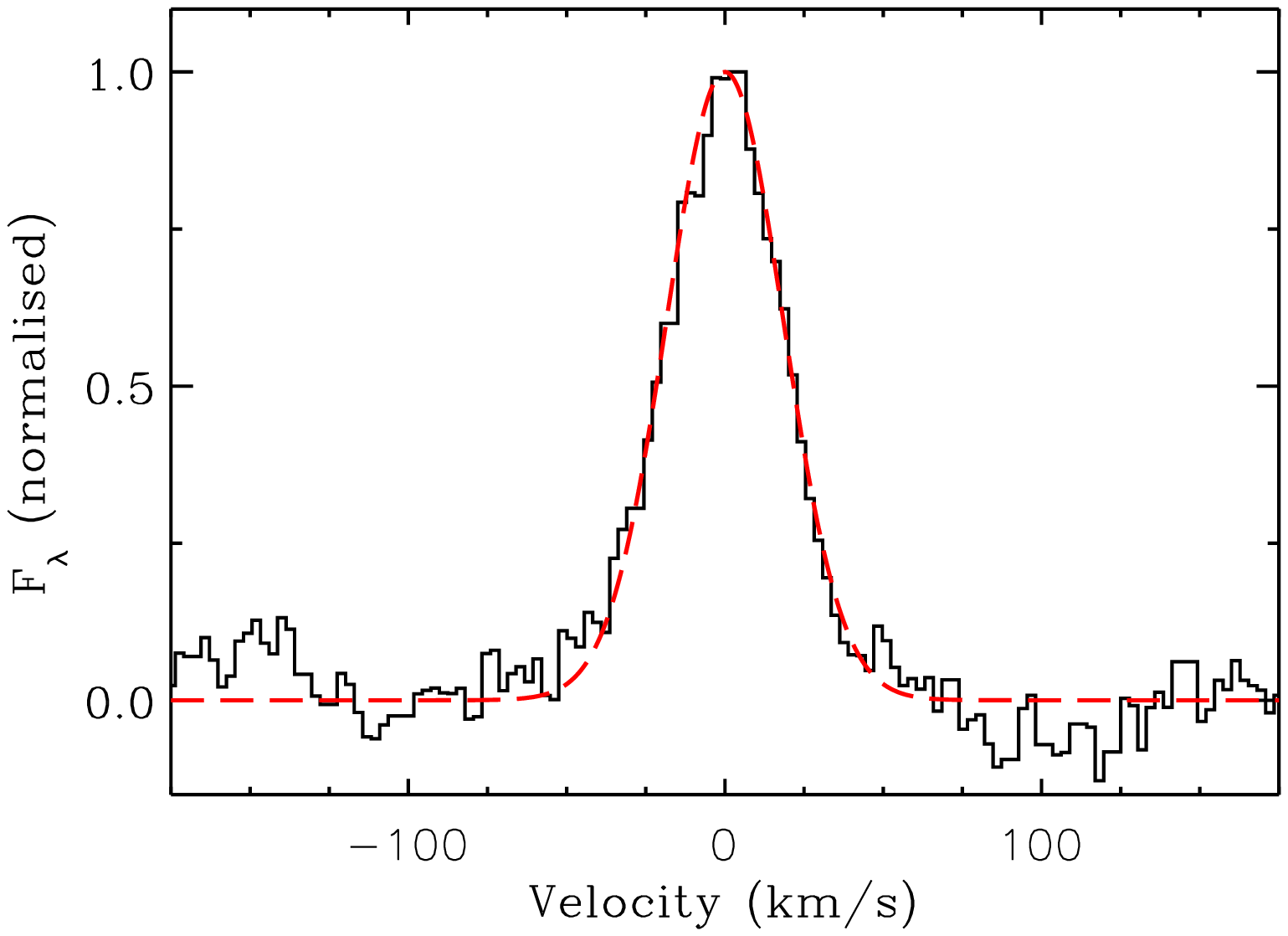}
\label{fig4e}}
\subfigure[MWC~349A]{
\includegraphics[width=2.7in,angle=0]{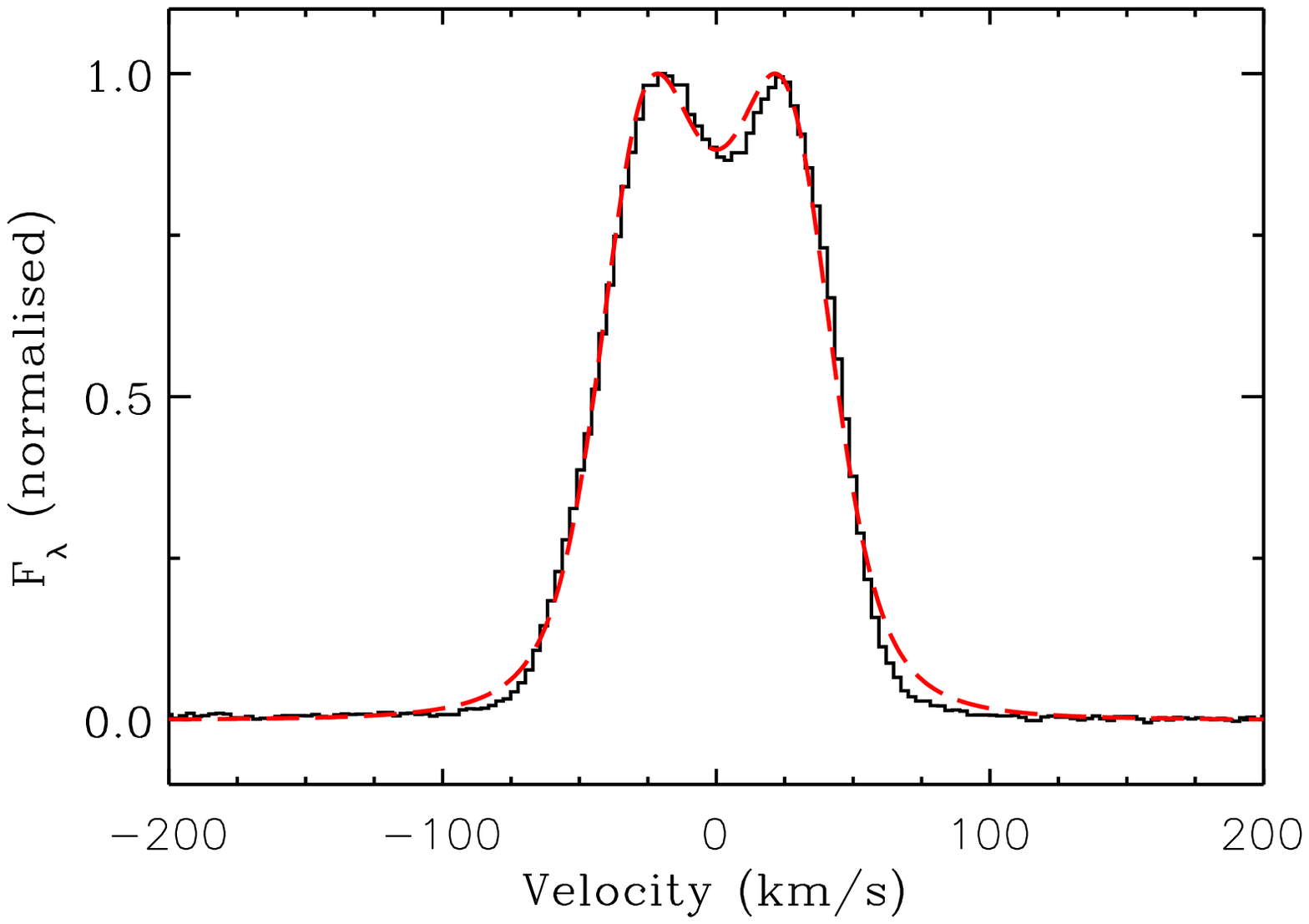}
\label{fig4f}}

\subfigure[S106 IRS]{
\includegraphics[width=2.7in,angle=0]{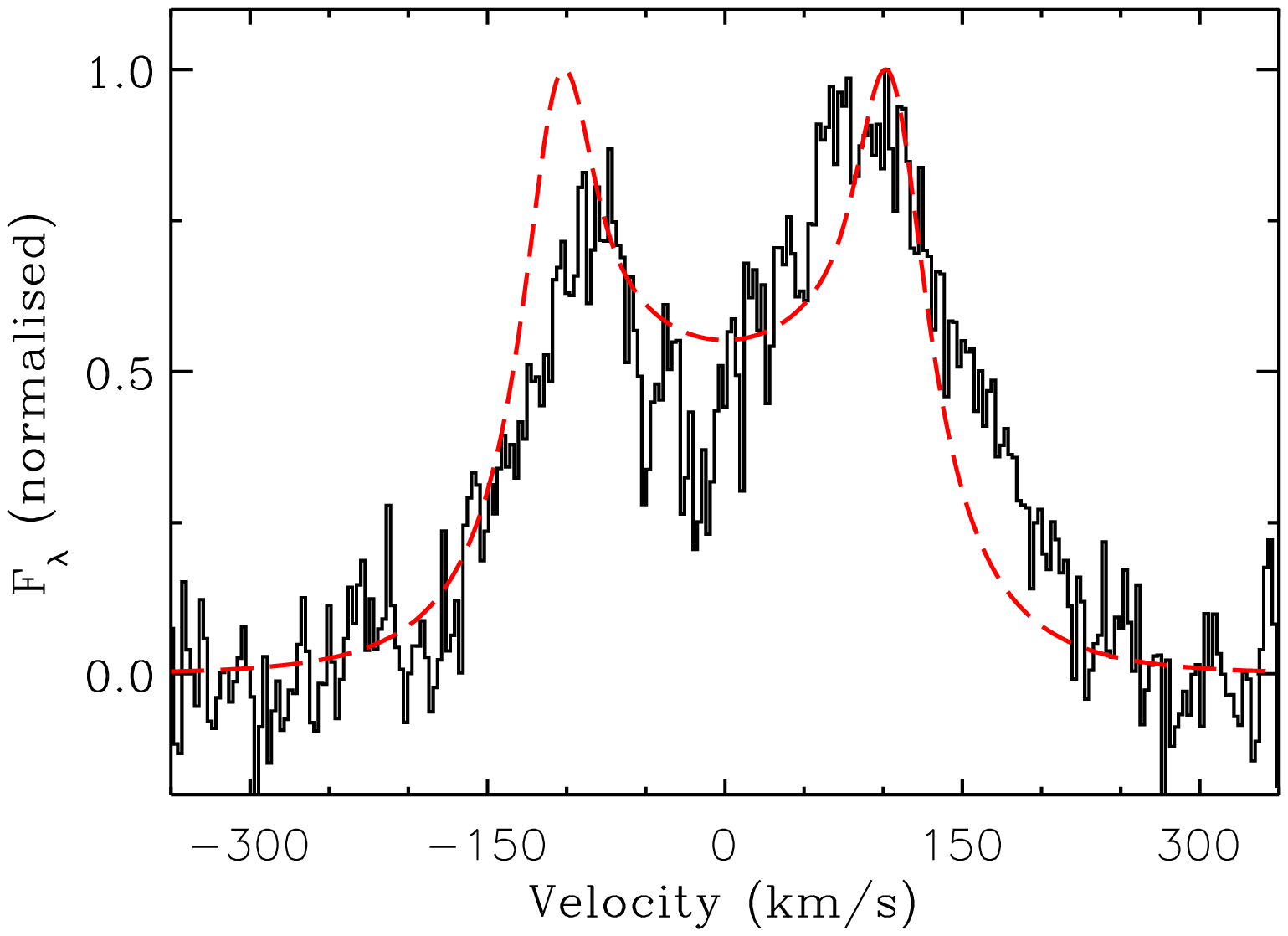}
\label{fig4g}}
\subfigure[VV Ser]{
\includegraphics[width=2.7in,angle=0]{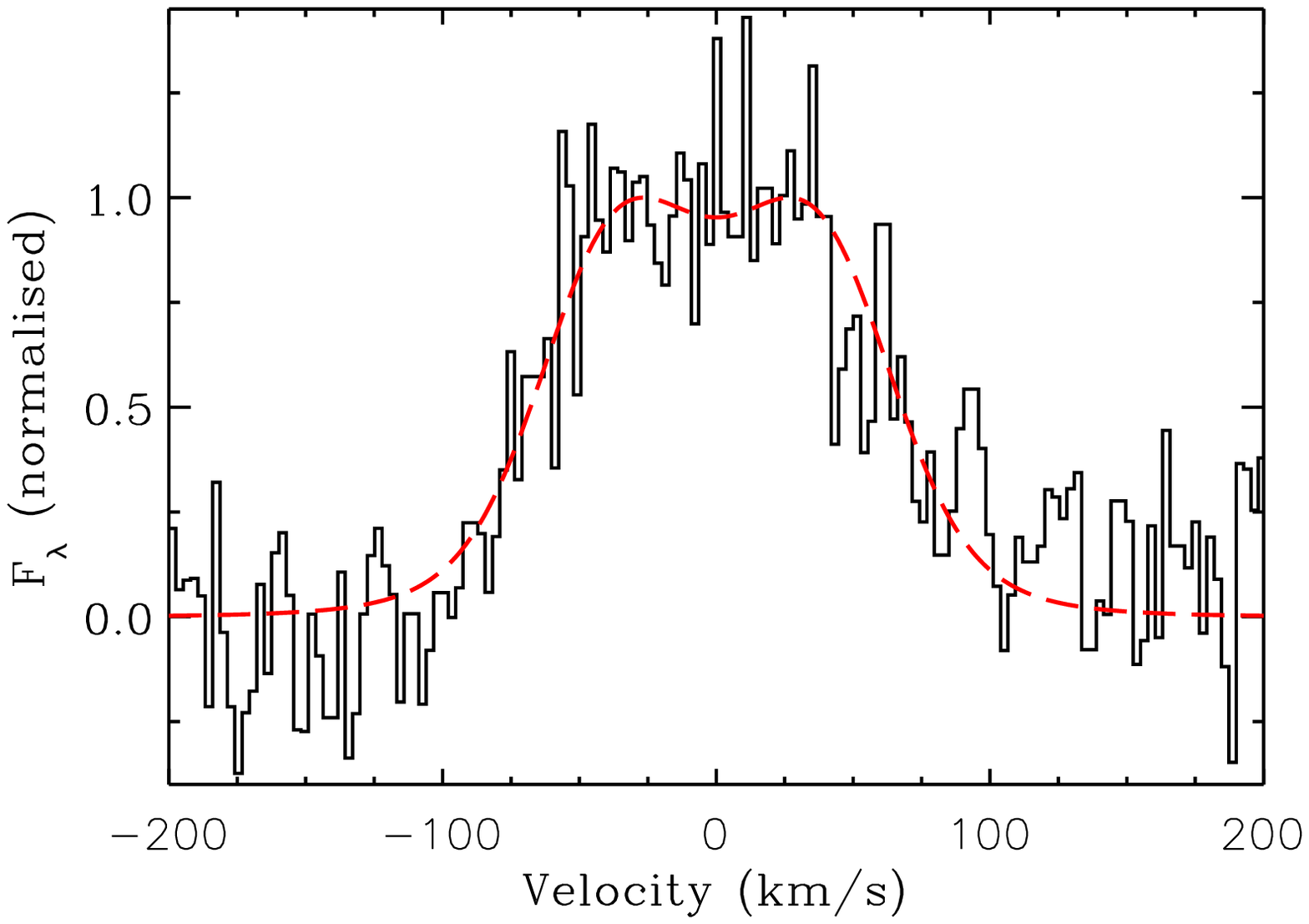}
\label{fig4h}}

\caption{FeII  emission line profiles of
  each object and the associated best-fitting disc model profiles (dashed line).  In 
VV Ser the ``spike'' evident in Figure \ref{fig2h} has been clipped
to allow a better disc fit to the underlying line profile.
\label{fe_ii_line_prof}}
\end{figure*}

\begin{table*}
\begin{tabular}{l l l l l l c l}
Object & $\mathrm{M_{\star}}$ &  $\mathrm{R_{\star}}$ & $i$ & $\mathrm{R_{out}}$ & Line-width & $\chi^2$\\
 & ($\mathrm{M_{\odot}}$) & ($\mathrm{R_{\odot}}$) & ($\mathrm{^{\circ}}$) & (AU) & ($\mathrm{km\,s^{-1}}$) \\

BD+40$^\circ$4124 & 6.1 & 3.56 & 45 & 21.6 &  60.8& 4.1\\

GL 490 & 8.6 & 4.28 & 30 &  5.2 &   75.0& 0.7 \\

M8E & 13.5 & 5.46 & 20 & 15.6&  44.6& 0.6\\

M17SW IRS1 & 15.0 & 5.80 & 15 &  3.7& 70.0& 0.7\\

M17SW IRS1 &  & &  67 & 67.5 &68.8 & 0.7\\

MWC~297 & 10.0 & 6.12 & 15 &  10.5 & 36.5& 0.5\\

MWC~349A & 26.0 & 8.54 & 85 &  31.0 & 26.4 & 41\\

S106 IRS & 21.7 & 7.51 & 80 &1.8  &  26.8& 1.0 \\

VV Ser & 3.8 & 2.81 & 85 &  2.0 &50.1  & 0.8\\

\end{tabular}
\caption{Parameters of the best fitting models.  The stellar radii and masses
  are taken from the compilation of Harmanec (1988).  For M17SW IRS1 we show
  two fits to the data, one reflecting the inclination inferred by Follert et
  al.\ (2010), the other for a more face-on orientation.  The fit shown in
  Figure \ref{fig4c} is the former, but there is essentially little difference
  between the two. The other inclinations are adopted from Table
  \ref{Properties}} \label{feii_fits}

\end{table*}

\subsection{Modelling of the FeII line with a disc}
The fluorescent FeII emission is a clear tracer of emission from a disk in
classical Be stars (Carciofi \& Bjorkman 2006).  Detailed models have been
derived for observations of the optical FeII lines in Be stars (eg Arias et al
2007), showing that even quite complex line profiles can be explained in terms
of emission from a relatively narrow ring.  It therefore seems natural to
consider whether a disc model can explain these relatively younger objects.
Currently, the 1.688$\mu$m FeII line is not yet included in any radiative
transfer model.  Nor do we understand its source function well enough, given
its potential origin as a laser (Johansson \& Letokhov 2007), to reliably model
even in the fashion of Arias et al.  We therefore rule out detailed
calculations at this point, and leave that to future work. Instead, we adopt a
``proof of concept approach'' and attempt to fit the observed line profiles
with a simple accretion disc model but without a detailed source function for
the line emission, or attempting to enforce a correct temperature structure for
the emission zone.  Specifically, we use the constraints given by other
observations to fix the mass of the star and the inclination (as quoted in
Table \ref{Properties}), and then test how well a disc can fit the observed
line profile.

The model in question consists of a Keplerian disc within which the line flux
per unit area is proportional to the local surface density. The structure of
the rotating disc is given by the standard $\alpha$-disc model -- as described
by, eg, Pringle (1981).  The accretion rate and value of $\alpha$ are not
fitted in this process, since the model is being fitted to a normalised line
profile, and thus the actual surface density is unimportant.  Furthermore, the
variation of surface density with radius is insensitive to both $\alpha$ and
the accretion rate. As a result, the only free parameters are the intrinsic
line width and the outer radius of the emitting area.  The latter is poorly
constrained for most of the objects however since they exhibit single peaked
emission.  In that instance the outer radius is a lower limit set by the model
line width, since we clearly cannot constrain this parameter strongly in the
case of a near face-on inclination.

Turbulent line broadening is added to the disc model in order to achieve the
final fits.  This is represented by a Gaussian, and the width is given in Table
\ref{feii_fits}.  No other line broadening mechanism is considered as we do not
expect any to occur -- even thermal broadening is likely to be insignificant
for iron.  Indeed, as noted by Johansson and Letokhov (2007), the conditions
under which this FeII line is emitted are likely to lead to mild lasing, which
actually narrows the natural line profile.  We also do not expect the opacity
in this transition to be significant.

The model is used to determine the line profiles for a given set of input
parameters. The model profiles are then smoothed to the instrumental
resolution, before being compared to the observed profiles. To find the best
fitting profiles, the outer radius and line width are allowed to vary between
1--100AU and 1--75kms$^{-1}$ respectively and the Powell method (as implemented
in {\sc{idl}}) is used to minimize the residual of the fit. We present a
comparison of the best fitting model profiles and the data in Figure
\ref{fe_ii_line_prof}. A detailed case-by-case discussion of the best-fitting
parameters is deferred to the next section, but it is clear that, in general,
the observed FeII 1.688$\mu$m line profiles can be modelled by emission from
circumstellar discs of an appropriate size.  The major exception to this is
S106 IRS, which we discuss in detail in Section \ref{s106comments}.  Figure
\ref{fig2} shows that GL~490, MWC~297 and MWC~349A also appear similar in FeII
and Br12 if we ignore the broad Br12 line wings.  A disc model can also partly
explain the Br12 emission in all of these sources as discussed below,
consistent with the results from fitting the iron line.  A disc model does not
fit the HI lines from the other sources well at all.

\subsection{Comments on individual objects}

\begin{figure*}
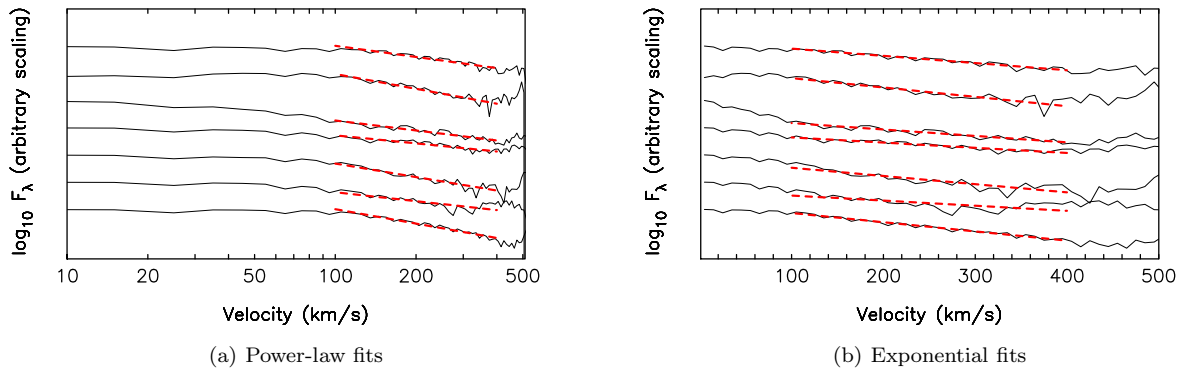

\centering
\subfigure[Power-law fits]{
\includegraphics[width=3.2in,angle=0]{fig_5a.ps}
\label{fig-f1}}
\subfigure[Exponential fits]{
\includegraphics[width=3.2in,angle=0]{fig_5b.ps}
\label{fig-f2}}
\caption{Fits to the Br12 line for either a power law or an exponential.  The
  fits are shown as the thick dashed (red) line.  The data have been binned
  into 10kms$^{-1}$ bins, and folded around the systemic velocity.  The fits
  are made only to the region between 100 and 400kms$^{-1}$.  The objects
  shown in each plot are (from top down): VV Ser, S106 IRS, MWC~297, M8E, M17SW
  IRS, GL~490 and BD+40$^\circ$4124.  }
\label{br12_fits}
\end{figure*}

\subsubsection{BD+40$^\circ$4124}


The H$\alpha$ line in this source shows an asymmetric double peak (Vink et
al.\ 2002), closer to line self-absorption than to evidence for a rotating
disk.  The velocity separation between the peaks is $\sim$120kms$^{-1}$,
compared to only about $\sim40$kms$^{-1}$ as seen in our Br$\gamma$ spectrum.
This is also consistent with the self-absorption picture since we would expect
H$\alpha$ to be more optically thick at velocities near, and to the blue of,
systemic compared to Br12 if a dense wind is present (as is evidenced by Figure
\ref{fig3a}).  Eisner et al.\ (2003) find an inner radius of 0.73au from
modelling their continuum interferometric data with a flat dusty disc.  Any
disc within this radius must be dust-free.  Vink et al.\ (2002) find that their
H$\alpha$ polarisation data can be explained by a straightforward line
depolarisation effect, where the continuum light is scattered by a small scale
disc, whereas the emission line itself arises on scales larger than the disc,
likely in a wind.  The small scale disc could be the extension of the dusty
disc detected by Eisner et al.\ down to the star.  The line wing/peak ratio is
roughly the same for Br$\gamma$ and H$\alpha$, but larger for Br12 (cf Figure
\ref{fig3a}), suggestive of Stark broadening.  The line wings can be fitted
equally well by both a power-law, with exponent $\sim-2.0\pm0.1$, and an
exponential (Figure \ref{br12_fits}).  We also explicitly fitted the Castor et
al (1970) model.  A formally good fit is achieved, but the key parameters in
this case, the width of the line core, the electron temperature (width) and
opacity, are highly degenerate.  Effectively there is little difference between
a very broad intrinsic line and a highly scattered one.  The wing--peak ratio
for the Br12 and Br$\gamma$ data, and the closeness of the power-law exponent
to the canonical $-2.5$, both suggest that Stark broadening is present however.
The FeII emission in this object is not fitted perfectly by a disc model, but
the profile is clearly very different from the Brackett lines.  In this
instance it is plausible that the hydrogen lines arise in a large scale wind,
whereas the FeII is mostly tracing the inner disk.

\subsubsection{GL 490}
Schreyer et al.\ (2006) detect a relatively large-scale ($\sim1500$au)
molecular disc in approximate Keplerian rotation from which they derive an
inclination of $\sim30^\circ$.  The position angle for this disc is the same as
that found at 2cm by Campbell, Persson \& McGregor (1986), and both lie
orthogonal to the direction of the observed large scale molecular outflow
observed by Mitchell et al.\ (1995), similar to other equatorial wind systems
(eg Hoare 2006).  Oudmaijer, Drew \& Vink (2005) also inferred the
presence of a small scale disc from near infrared spectropolarimetry.
Oudmaijer et al.\ did however require the initial line emitting material to lie
internal to the scattering disc, and that the disc did not extend down to the
stellar surface.

Our Br12 profile looks very similar to the Br$\gamma$ profile in Bunn et
al.\ (1995).  Bunn et al.\ also infer a relatively flat Br$\alpha$/Br$\gamma$
ratio in GL~490 out to velocities $\sim\pm100$kms$^{-1}$, before an upturn is
seen.  However, the evidence for a turn-up in these earlier CGS4 data may be
compromised by poorer continuum subtraction, since the data were taken with a
smaller array and at lower spectral resolution, spatial resolution and free
spectral range.  Comparison of our Br12 profile with the Br$\gamma$ and
Br$\alpha$ profiles in Bunn et al.\ suggests that the broad line wings are
relatively stronger in Br12 which slightly favours Stark broadening.  There is
a good fit to the Castor et al model for Br 12 as well, and in this instance
the parameters are well defined with $\tau\sim0.3\pm0.05$ and
$T_e\sim340\pm30$K.  The very low value of the electron temperature implied by
the fit is a concern, but we do not have sufficient signal-to-noise to
determine if the wings extend further in velocity.  The line wings do not
follow the expected power law for Stark broadening either, having an exponent
of only $\sim-1.2\pm0.2$. In this instance, the relatively poor signal-to-noise
prevents a truly firm conclusion.

The Brackett lines differ from the FeII line largely in a strong red asymmetry,
and stronger evidence for broad line wings in the hydrogen lines.  The
FeII emission can be successfully fitted with a Keplerian disc.  The inferred
lower limit to the outer radius is much smaller than the size of the structure
seen in the radio map which is several hundred AU across.  Given the similarity
of the line core in this object between Br12 and FeII it is possible that a
disc may be a suitable explanation for the hydrogen line emission as well,
especially given the orientation of the radio emission.  There is no
self-absorption in any of the lines observed either here or in Bunn et al.\ so
there is no requirement for a strict wind component to be present.

\subsubsection{M17SW IRS1}

M17SW IRS1 is a young binary star located just beyond the edge of the M17 HII
region and is also known as the Kleinmann-Wright object (Kleinmann \& Wright
1973).  Estimates of its luminosity range from about 5000\Lsolar\ (Chini et
al.\ 2004) to 50000\Lsolar\ (Follert et al. 2010).  The latter is more
consistent with the presence of HeI 2.058$\mu$m emission as reported by both
Porter et al.\ (1998) and Hanson et al.\ (1997).

The deredenned flux ratio is very similar to that presented by Bunn, Drew \&
Hoare, and as noted by them is rather similar to that for BD+40$^\circ$4124.
 Br$\gamma$ does show weak red asymmetric self-absorption
which might suggest the presence of infalling ionised gas.

The Br12 and FeII emission have very different profiles, again in a similar
fashion to BD+40$^\circ$4124.  The FeII data can be explained by a disc, at
both a near edge-on inclination as measured by Follert et al.\ (2010), as well
as nearer to face-on (cf Table \ref{feii_fits}).  However, an edge-on disc
requires a very large spread in the FeII emission region, almost out to the
full extent of the mid infrared emission region seen by Follert et al.  Without
further information however we cannot rule out either possibility.  The
degeneracy between inclination and outer radius of disc is of course a well
known problem in this situation, but M17SW IRS1 is the only source in which the
result of fixing the inclination at the value given in Table \ref{Properties}
leads to results that are then difficult to interpret in light of other data.
The most likely model for this source is the same as discussed for
BD+40$^\circ$4124, namely a disc giving rise to the FeII emission and a wind
the Brackett line emission.  The wings on Br12 are relatively weak in this
source, but the power law profile of the wing is $\sim-1.8\pm0.2$, suggesting
that Stark broadening may be present.  A formally good fit to the Castor et al
model can be achieved but the values of both opacity and electron temperature 
are again highly degenerate with the core line width, and no meaningful
physical parameters can be obtained.

\subsubsection{M8E}

The Br12 line in M8E is best described as a relatively narrow Gaussian
superposed on a much broader ``pedestal''.  This is qualitatively different
from the Br$\alpha$ and Br$\gamma$ profiles in Bunn, Drew \& Hoare.  The line
core is similar to Br$\gamma$, but the higher velocity line wings are less
evident.  Br$\alpha$ has broad wings but the profiles are not a match.  It is
possible the greater wavelength coverage of the data presented here may
partially explain this difference.  It cannot be explained easily with simple
line broadening mechanisms outlined in Section \ref{hydrogen} since there is no
simple trend with quantum number $n$. An exponential fit is a better match to
the broad wings than a power-law, and the latter also has a very low exponent
of $-1.1\pm0.1$.  Neither matches at the highest velocities
however.  The Castor et al model is formally a very poor fit.

One other notable aspect of our data is that the line centre velocity of the
Brackett lines are significantly offset with respect to systemic.  As shown by
Table \ref{SimpleFits}, these lines have $v_{LSR}\sim-37$kms$^{-1}$, whereas
the FeII line has $v_{LSR}\sim9$kms$^{-1}$, which is in agreement with the
measured molecular gas value, and the Br$\alpha$ and Br$\gamma$ data of Bunn et
al.\ (1995).  The fact that both Br11 and Br12 lines agree in velocity rules
out a simple wavelength error in part of the data since they are observed in
different echelle settings.  Bunn et al.\ infer a relatively flat
Br$\alpha$/Br$\gamma$ optically thick ratio in M8E out to velocities
$\sim\pm100$kms$^{-1}$, with only weak, if any, evidence of a turn-up at higher
velocity.

The FeII line can be modelled by a disc, but the relatively weak line emission
is noisy and hence does not particularly constrain other models.  Linz et
al.\ (2009) can only set an upper limit on the size of the disc in the
mid-infrared of $\sim$100AU, which is consistent with our FeII disc model.  The
similarity of the core of the Br12 line and the FeII line, if we ignore the
velocity offset mentioned above, suggests both have a common origin.  Overall
though the difference in velocity, the unusual ``pedestal'' on Br12 and the
fact that the line widths do not simply scale with $n$ all make it difficult to
ascribe an origin to the emission.

\subsubsection{MWC~297}\label{m297comments}

\begin{figure*}
\begin{center}
 \includegraphics[width=0.6\textwidth,angle=0]{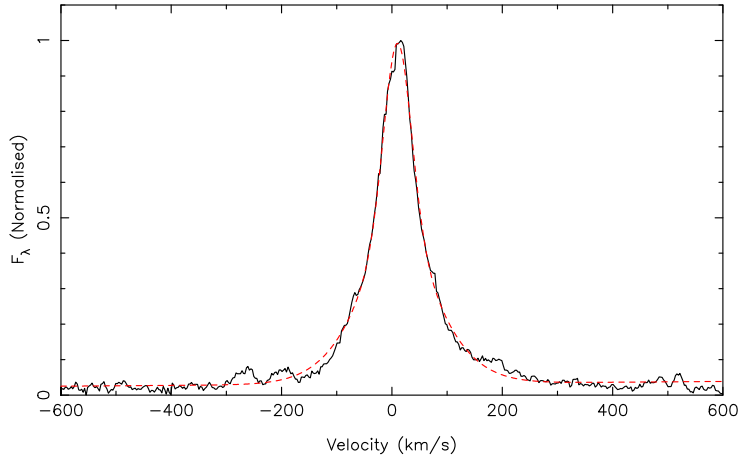}
  \caption{The best fitting Castor et al.\ (1970) model to MWC297.  
}
  \label{mwc297-fit}
\end{center}
\end{figure*}

The inclination of this source is the subject of debate.  Drew et al.\ (1997)
find a near edge-on inclination is necessary to explain the observed optical
absorption line data, since otherwise the star is near break-up velocity.  By
contrast all of the interferometry data is best modelled by a near face-on
orientation instead.  An edge-on model is also more likely to lead to
double-peaked line profiles if an ionised disc is a contributor.  For
consistency with the other objects studied here, where we have adopted
inclinations from interferometric data, we do the same here.

The Br$\alpha$/Br$\gamma$ ratio presented in Murdoch \& Drew (1994) is similar
to the Br$\gamma$/Br12 ratio of BD+40$^\circ$4124 or M17SW IRS1 shown in Figure
\ref{fig3}.  The line widths we measure are similar to the Br$\alpha$ width as
reported in Drew et al.\ (1997).  The FWZI of Br12 is closer to that seen in
H$\alpha$, though, as is clear from Drew et al., the latter has a much larger
FWHM.  The FeII profile does not show the same broad wings.  Most
published spectra of MWC~297 show symmetric Brackett line profiles.  The
exception is Malbet et al.\ (2007), where a clear blue-shifted asymmetric
self-absorption is present in Br$\gamma$.  The higher spectral resolution data
presented by Murdoch \& Drew (1994) or Weigelt et al.\ (2011) do not show this,
which suggests variability is present.  There is a clear P-Cygni effect seen in
the 1.083$\mu$m 2$^3$S--2$^3$P HeI line (Drew et al.\ 1997), suggesting a wind
is present, which for a relatively face-on orientation must either be a polar
wind or a disc-wind with a large opening angle.  

Drew et al.\ present a 5GHz radio map showing resolved radio emission from this
source on the scale of $\sim50AU$, and both the Malbet et al.\ and Weigelt et
al.\ interferometry also show the core Br$\gamma$ emission is more extended
than the K-band continuum light.  The higher spectral resolution data of
Weigelt et al.\ show the visibility is essentially the same as the continuum
beyond about 60kms$^{-1}$ from the line centre.  The Br12 wings is better
fitted by an exponential in this case.  Notably this is the only object in
which the line core clearly has greater flux than a simple extrapolation of the
exponential fit might suggest (Figure \ref{fig-f2}), as we would expect if the
exponential wings were light scattered from that core.  A power-law fit by
comparison again has a rather low exponent of $-1.3\pm0.1$, and the flat inner
line core from Stark broadening seen in the Repolust et al.\ data, and objects
such as BD+40$^\circ$4124, is absent here.  Comparison of the strength of the
line wings to peak intensity in the hydrogen recombination lines in the data
presented here and in Drew et al.\ (1997) indicates that the wings grow
stronger for the lower $n$ transitions in a fairly uniform way.  This is
strongly suggestive of electron scattering.  The fact that the wings and
continuum appear to have the same visibility in the Weigelt et al.\ data
suggests this scattering occurs on small scales.

We consider the scattering model in more detail for MWC~297 since this source
shows the best evidence that it is present.  The best fit of the Castor et al
model has a reduced $\chi^2=2$, indicative of the fact that the line is
slightly asymmetric.  We show the fit for this source in Figure
\ref{mwc297-fit}.  In this case the opacity to Thomson scattering is greater
than unity, so the conditions on the actual model are not met.  The implied
electron temperature is also very low ($T_e\sim170\pm10$K) and not physically
meaningful.  The match of the Castor et al profile in a case where the
intrinsic line core itself is very narrow does argue for scattering being the
dominant mechanism, but that this model itself is insufficient to accurately
estimate key physical parameters related to the scattering.

If we accept a mostly face-on inclination for this source then our FeII profile
can be fitted well by a disc model, with an outer radius that is well within
the extent of the ionised gas as seen in the radio map.  Acke et al.\ (2008)
struggled to fit their combined near and mid-infrared interferometry with a
pure accretion disc model, suggesting that more than one component was present.
Malbet et al.\ suggest a combination of disc and a weak outflowing wind can
explain their near infrared interferometry and the observed P-Cygni profile
they find in Br$\gamma$, but they struggle to find a good fit to the red wing
of the Br$\gamma$ line because their model predicts disc shadowing should be
present where none is seen.  The same is true of the models applied to the high
spectral resolution observations of Weigelt et al.\ (2011) as well (which
otherwise reach a very similar conclusion to Malbet et al).  However the
relatively clear evidence for significant scattering in the line wings largely
reduces this problem.  In a face-on orientation, the lack of a clear line
effect in the spectropolarimetry of Oudmaijer \& Drew (1999) is not
significant, since an asymmetry in the source structure as seen on the sky is
required to give rise to such a signature, and this is naturally not present.
A counter argument to this is that Weigelt et al.\ find evidence from the
differential phase for a rotational signature in Br$\gamma$, which of course we
would not see in an almost completely face-on situation.  An intermediate
inclination, as suggested by Acke et al.\ (2008) as a compromise, may help to
reconcile all of these pieces of evidence.

\subsubsection{MWC~349A}\label{m349comments}

MWC~349A clearly shows different line profiles and ratios to our other sources.
The precise evolutionary status of MWC~349A is not known (see, eg, Hofmann et
al.\ 2002 for a discussion), nor is an exact spectral type known.  The main
argument against it being young is that it lacks significant far-infrared
thermal dust emission, with the dust emission peaking in the mid-infrared
instead.  The strong emission line spectrum and bolometric luminosity are also
consistent with an identification as an evolved B[e] supergiant, though it also
lies at the upper end of properties of known Herbig Be stars.  The luminosity
is $6\times10^4$\Lsolar (adapted from Cohen et al.\ 1985 using the distance
from Meyer et al.\ 2002).  This is consistent with the presence of HeI lines in
the spectrum (eg Hamann \& Simon 1986).  MWC~349A shows strong hydrogen maser
recombination lines in the mm and radio regime (eg Weintroub et al.\ 2008),
consistent with the edge-on inclination found from interferometry.

FeII emission from similar energy levels to the 1.688$\mu$m line also show
double peaked profiles (eg the 2.089$\mu$m FeII line in Hamann \& Simon 1986,
the optical lines in Hamann \& Simon 1988).  These lines have been interpreted
as having an origin in a disc (Hamann \& Simon 1988).  Higher excitation FeII
lines appear more like the Br$\gamma$ profile we see here, with a distinct line
asymmetry (Hamann \& Simon 1988).  MWC~349A is the object where our data have
the highest signal-to-noise, and therefore small departures from a Keplerian
disc are easier to identify.  This is clear in our FeII disc fit, which
formally is very poor.  However, this is mostly due to a weak blue-red
asymmetry in the observed data, which cannot be modelled within the context of
our very simple prescription.  Kraus et al.\ (2000) also had some difficulty in
fitting the v=2--0 CO bandhead with a Keplerian disc profile.  A disc is
clearly present in this source, however, given the maser emission.  It may
simply be that the disc is not completely Keplerian as also suggested by the
maser observations of Weintroub et al.\ (2008).  Danchi et al.\ (2001) derived
an outer radius of about 50AU from near-infrared interferometry, consistent
with what we find from modelling the FeII data, and the origin of the mm/sub-mm
masers (Thum et al.\ 1994), though all are rather larger discs than the best
fit that Kraus et al.\ find for their disc model.

The HI Br$\gamma$ profile also appears very similar to that shown in Hamann and
Simon (1986), and shows what is likely to be weak blue self-absorption or a
very asymmetric disk.  The broad wings can again be ascribed to
electron scattering (Meyer et al.\ 2002), in agreement with the
spectropolarimetry of Oudmaijer et al.\ (2005).  The hydrogen lines have
generally been interpreted as the combination of emission from a slow stellar
wind and a rotating disc (eg Hamann \& Simon, 1986, 1988).  This wind must be
on larger scales than the disc modelled by Danchi et al., since otherwise the
velocities would be anomalously low.  Another possibility is a disc-wind since
the low expansion velocity evident in the hydrogen lines naturally arises in a
model where the wind is launched from the disc at some distance from the star
(eg Gordon 2003, Hollenbach et al 1994, Sim et al.\ 2005).

The Br$\gamma$/Br12 line ratio plot shows a distinctive pattern different from
any of either our sources or those presented in Bunn et al.  It is however
consistent with self-absorption in the lines due to an outflowing wind in the
sense that the opacity increases uniformly towards the blue.  We also
considered the ratio after the individual line profiles had been
cross-correlated, in order to compensate for any residual velocity calibration
errors that might be present, which could have an effect here given the
steepness of the profile.  The result shows a slightly flatter line ratio, but
still with a distinct slope in the same sense as shown here.  Kraus et
al.\ found that the high series Pfund lines were singly peaked with widths
$\sim50$kms$^{-1}$.  This suggests these lines arise in optically thin
gas far from the star in the disc-wind, or in the nebula.  However such a large
scale optically thin Pfund component should be reflected in both the Balmer and
Brackett lines as well, and is not. The mid infrared interferometry presented
by Quirrenbach, Albrecht \& Tubbs (2006) also shows the high excitation mid
infrared hydrogen lines are less extended than the dust continuum emission.
Instead the Pfund and higher series lines must, on energetic grounds, arise
near the central source, and the single peaked nature implies optically thin,
approximately spherical, wind emission.  It is probable therefore that the
infrared recombination lines largely arise from a combination of disc-wind and
spherical wind.

\subsubsection{S106 IRS}\label{s106comments}

S106 IRS is the exciting star of the Sh 2-106 HII region (see, eg, Saito et
al.\ 2009).  The HII region is best modelled as a bipolar flow which is almost
oriented in the plane of the sky (eg Solf \& Carsenty 1982).  S106 IRS has a
variety of reported extinction values in the literature.  The value given in
Table \ref{Properties} is from a fit to the silicate absorption spectrum as
measured by ISO in a large beam (van den Ancker et al.\ 2000) which will be
dominated by the emission from S106~IRS.  The fits to the CO bandhead by
Chandler et al.\ (1995) shows that a disc of radius 1--2AU is likely to be
present.

\begin{figure*}
\begin{center}
 \includegraphics[width=0.6\textwidth,angle=0]{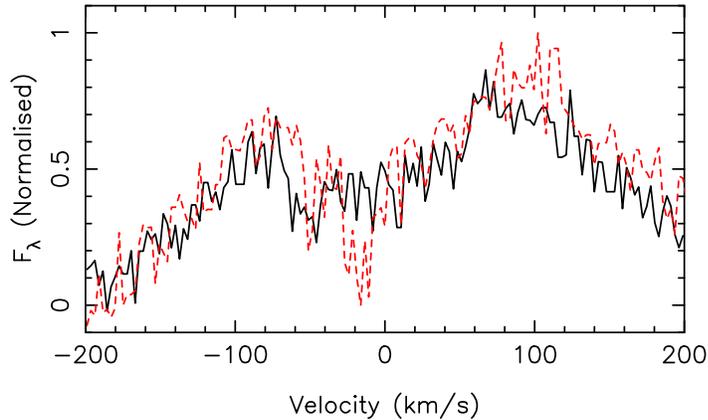}
  \caption{The profiles  of the FeII 1.688$\mu$m line in S106 IRS
separated by 378 days.  The first epoch is the solid line.}
  \label{S106-epoch}
\end{center}
\end{figure*}

The Br$\gamma$ and FeII lines in S106 IRS are remarkably similar (Figure
\ref{fig2g}), though Br$\gamma$ has a larger FWZI than the FeII line.  Perhaps
the greatest surprise is that despite the near edge-on nature of this source, a
disc model is a very poor fit (Figure \ref{fe_ii_line_prof}).  We observed the
FeII line in S106 IRS at two epochs separated by 378 days.  The data shown in
Figure \ref{fig2g} are the average of these observations.  There was no
evidence that the HI Br11 line varied between these observations but there is
some tentative evidence of a change in the FeII line.  This is shown in Figure
\ref{S106-epoch} where we plot the data from the two epochs, after slight
smoothing to improve the signal-to-noise.  There is a clear dip in the data at
the velocity of maximum absorption for the second epoch.  We can understand
both the line profile and the variability if the FeII line in this case does
arise in a clumpy (hence variable) wind.  The remaining differences in the line
profile between the Br12 and FeII line then are due largely to the differing
opacities in the two lines.  We already know that S106 IRS shows strong P-Cygni
absorption in the HeI 2$^1$S--2$^1$P transition at 2.058$\mu$m (Drew et al.\
1993), indicative of a stellar wind with a high mass loss rate (see also Felli
et al.\ 1984).

%

The Br$\gamma$/Br12 ratio plot for this source is similar to the
Br$\alpha$/Br$\gamma$ ratio presented in Drew et al.\ (1993) but with better
signal-to-noise at high velocity.  Drew et al.\ found an upturn at
$v_{LSR}\sim-200$kms$^{-1}$ in their Br$\alpha$/Br$\gamma$ ratio, which they
interpret as evidence that Br$\gamma$ is becoming optically thin.  However we
see no such evidence.  Drew et al.\ tried to deblend the prominent
5$^{1,3}$G-4$^{1,2}$D HeI line from Br$\alpha$ at around this wavelength,
whereas we have not done the same for the 2.16475$\mu$m 7$^{3,1}$G--4$^{3,1}$F
HeI lines which may explain the difference, though it should in theory enhance
the up-turn in our data and suppress it in theirs.  We therefore accept our
line ratio as indicative for the source as a whole.  A broadening mechanism is
therefore required to explain the line wings as in other sources.  The Br12
wings have a power law profile with exponent $-2.0\pm0.2$, similar to the case
of BD+40$^\circ$4124.  An exponential or the Castor et al model give a slightly
poorer fit.  Since our Castor et al model does not match the central absorption
we do not consider the fitting parameters further for this source.  The
enhanced wings seen in the Brackett lines appear more prominently in Br12 than
Br$\alpha$ however, more suggestive of Stark broadening.

An almost unresolved nebular core further complicates the picture for our
Br$\gamma$/Br12 ratio.  The nebular line arises from a much larger volume than
the rest of the emission, and should have a lower extinction correction applied
since the HII region is optically visible.  This explains why the ratio plotted
in Figure \ref{fig3d} exceeds the nominal case B value.

Overall most of the lines in this source can be explained by a dense wind (as
in Drew et al.\ 2003) which must be primarily along the equatorial direction to
avoid veiling the nebula from the star.  In this instance, the FeII emission
may also arise from the disc-wind, given the asymmetric wind-like profile it
has.

\subsubsection{VV Ser}

Blake \& Boogert (2004) observed the M band fundamental CO transition in this
source.  Their results show a clear double peaked profile which they model as
a disc with $i=85^\circ$, consistent with the interferometry of Eisner et
al.\ (2003).  Notably the blue peak of their CO profile is enhanced with a
narrow emission component in the same fashion as our FeII profile.  This is
suggestive of a localised spot of enhanced emission, which must have an origin
in a source with constant velocity (ie not the disc), since our data and the
Blake \& Boogert data are separated by 2--3 years. We do not quite
resolve the blue and red peak in this source despite the almost edge-on nature.
The disc model fits well despite this if we ignore the extra ``spike''.
Notably, Eisner et al.\ (2003) suggest that a disc of similarly small size is
present in their interferometry, though they favour one with a puffed up inner
geometry.  Pontopiddan, Blake \& Smette (2011) model the CO as arising at
3.5au, outside the emission region for the $K$-band continuum.  VV Ser is
observed to have a disk shadow, consistent with this modelling (see discussion
in Pontopiddan et al.\ 2011).

The H$\alpha$ profile is asymmetric double peaked (Reipurth, Pedrosa \& Lago,
1996) with a velocity separation of 250kms$^{-1}$ between the red and blue
peaks, again suggestive of a weak P-Cygni profile.  Our data lack such obvious
P-Cygni effects, but the ratio shown in Figure \ref{fig3e} is otherwise similar
to most of the other objects.  An exponential fit to the line wings is
marginally preferred.  A power-law fit gives an exponent that is again rather
shallow at $-1.5\pm0.1$.  The line wings are stronger relative to the peak in
the lower series lines for VV Ser (eg Figure 3(e)), which is suggestive of a
scattering origin as well.  The Castor et al model gives a formally good fit
but the opacity and electron temperature are effectively undefined in the
result as was the case for other sources due to degeneracy in the best fitting
model.  A larger wavelength coverage would perhaps have helped identify the
source of the broad line wings in this source.  A wind origin for the remaining
hydrogen emission seems likely given the very different Br12 and FeII profiles,
with the latter almost certainly coming from a disc (Figure \ref{fig2h}).

\section{Conclusions}

Our results show clearly that a wide variety of physical environments can be
traced using near infrared spectral lines.  It is clear that the combination of
HI, FeII and, where available, HeI lines can help to disentangle the otherwise
complex line profiles observed.  Hydrogen line ratios are particularly powerful
in showing the need for a line broadening mechanism in all objects expect
MWC~349.  Specifically, data with sufficient spectral resolution {\em and}
wavelength coverage allowed us to identify a likely origin in Stark broadening
in the wings of hydrogen lines in BD+40$^\circ$4124, M17SW IRS and S106 IRS,
and electron scattered wings in MWC~297, and probably VV Ser.  Better
signal-to-noise, as available with larger telescopes, and an even larger free
spectral range would have allowed us to be more conclusive about the origin of
the wings in both GL490 and M8E as well.

One important consequence of the presence of strong line broadening mechanisms
in the high series infrared hydrogen lines is that, on their own, they are poor
tracers of a wind or a disc.  A better tracer of a wind are undoubtedly the HeI
lines from metastable levels, although, unfortunately, not every young stellar
object shows this emission. The hydrogen lines do however provide a means to
test for the presence of very dense circumstellar material (through
broadening).

We also attempted to utilise the model of Castor et al.\ (1970) to estimate
physical scattering parameters, assuming that was the dominant mechanism.  The
fits were either highly degenerate in the parameters where the core line width
is close to matching that required for the wings (eg BD+40$^\circ$4124, M17SW
IRS and VV Ser), or else returned values that indicate that although the
profile can be a good match, the estimated parameters are not physically
sensible (eg MWC~297).  It may be that the success of this model in the case of
Lk H$\alpha$ 101 by Hamman and Persson (1989) was fortuitous, since it is a
very similar type of object.  They fitted to H$\alpha$ which will have more
easily detectable wings for electron scattering, and hence they can be seen to
higher velocities which is what drives the fit to the electron temperature.  It
is worth considering briefly why the model fits at all if the parameter
estimation is so poor.  The reason is simply that outlined by Laor (2006) --
multiple scattering gives rise to a close to exponential profile, and the high
velocity component of the Castor et al model is essentially very close to an
exponential.  It seems clear that a full Monte-Carlo approach to modelling
these systems, such as that outlined in Kurosawa et al.\ (2011), is required to
proceed much further with any analysis.

We have also shown that the 1.688$\mu$m z$^4$F$_9$--c$^4$F$_9$ FeII line is
seen in massive and intermediate mass young stellar objects.  We have
successfully modelled this line with a simple Keplerian disc in all sources bar
S106 IRS.  The model results are largely in agreement with those found from
interferometric near and mid infrared measurements of the continuum where such
data exist.  We therefore ascribe a disc origin to the FeII line in most
sources, since we know from the interferometry and other data that these
sources do have discs, and our simple model fits are largely successful.  This
is in agreement with previous findings from optical data that fluorescent FeII
emission is a good disc tracer in classical Be stars.  The counter-example of
S106~IRS cautions against too simple an over-interpretation of this result
however.

Overall therefore, these data are strongly suggestive that all of the sources
observed have evidence that a disc is present, as expected for such young
sources, and the line broadening indicates a very dense circumstellar
environment.  The likeliest origin for the bulk of the emission lines is a
combination of a disc and a wide opening angle wind or disc-wind.  The relative
contributions of the latter varies widely between different species and even
different lines of the same species.  The similarity of features between these
sources and lower mass T Tauri stars is notable.

\section{Acknowledgments}

We are grateful to the referee Juan Zorec for his very helpful referee's
report. 
The United Kingdom Infrared Telescope is operated by the Joint Astronomy Centre
on behalf of the Science and Technology Facilities Council of the U.K.  HEW
acknowledges support from STFC through a studentship and as a postdoctoral
research associate.

\parindent=0pt

\vspace*{3mm}

\section*{References}
\begin{refs}
\mnref{Acke, B. 
  et al., 2008, \aap, 485, 209}
\mnref{Alonso-Albi, T., Fuente, A., Bachiller, R., Neri, R., Planesas, P.,
  Testi, L., Bern{\'e}, O., Joblin, C., 2009, \aap, 497, 117}
\mnref{Arias M.L., Zorec J., Cidale L., Ringuelet A.E., Morrell N.I., 
Ballereau D., 2006, A\&A, 460, 821 }

\mnref{Baldwin, J.A., Ferland, G.J., Korista, K.T., Hamann, F., 
   LaCluyz{\'e}, A., 2004, \apj, 615, 610 }
\mnref{Bik, A., Thi, W.F., 2004, \aap, 427, L13 }
\mnref{Blake, G.A., Boogert, A.C.A., 2004, \apj, 606, L73} 
\mnref{Blum, R.D., Barbosa, C.L., Damineli, A., Conti, P.S., 
    Ridgway, S., 2004, \apj, 617, 1167 }
\mnref{Bonnell, I.A., Vine, S.G., Bate, M.R., 2004, \mnras, 349, 735 }
\mnref{Bunn, J.C., Hoare, M.G., Drew, J.E., 1995, \mnras, 272, 346 }
\mnref{Campbell, B., Persson, S.E., McGregor, P.J., 1986, \apj, 305, 336 }
\mnref{Canto, J., Rodriguez, L.F., Calvet, N.,  Levreault, R.M.,
  1984, \apj, 282, 631 }
\mnref{Carciofi, A.C., Bjorkman, J.E., 2006, ApJ, 639, 1081}
\mnref{Carciofi, A.C., Domiciano de Souza, A., Magalh{\~a}es, A.M.,
	Bjorkman, J.E., Vakili, F., 2008, ApJ, L41}

\mnref{Cardelli, J.A., Clayton, G.C.,  Mathis, J.S., 1989, \apj, 345, 245 }
\mnref{Carr, J.S., 1989, \apj, 345, 522 }
\mnref{Castor, J.I., Smith, L.F., van Blerkom, D., 1970, 159, 1119}
\mnref{Chandler, C.J., Carlstrom, J.E., Scoville, N.Z., 1995, \apj, 446, 793}
\mnref{Chini, R., Hoffmeister, V.H., K{\"a}mpgen, K., Kimeswenger, S.,
  Nielbock, M.,    Siebenmorgen, R., 2004, \aap, 427, 849 }
\mnref{Cohen, M., Bieging, J.H., Welch, W.J., Dreher, J.W.,
  1985, \apj, 292, 249 }
\mnref{Corcoran, M., Ray, T.P., 1997, \aap, 321, 189 }
\mnref{Crowther, P.A., 2005, in Cesaroni, R., Felli, M., Churchwell, E.,
  Walmsley, M., eds, IAU Symp.\ 227, 
  ``Massive Star Birth: A Crossroads of
   Astrophysics'', p.\ 389
  }
   
\mnref{Danchi, W.C., Tuthill, P.G., Monnier, J.D., 2001, \apj, 562, 440 }
\mnref{Davies B., Lumsden S.L., Hoare M.G., Oudmaijer R.D., de Wit W.-J.,
  2010, MNRAS, 402, 1504} 
\mnref{de Wit W.J., Hoare M.G., Oudmaijer R.D., N{\"u}rnberger D.E.A., 
Wheelwright H.E., Lumsden S.L., 2011, A\&A, 526, L5 }
\mnref{Drew, J.E., Bunn, J.C., Hoare, M.G., 1993, \mnras, 265, 12 }
\mnref{Drew, J.E., Busfield, G., Hoare, M.G., Murdoch, K.A., Nixon, C.A., 
 Oudmaijer, R.D., 1997, \mnras, 286, 538 }
\mnref{Edwards, S., Fischer, W., Hillenbrand, L.,  Kwan, J.\ 
  2006, \apj, 646, 319}
\mnref{Eisner, J.A., Lane, B.F., Akeson, R.L., Hillenbrand, L.A., 
  Sargent, A.I., 2003, \apj, 588, 360 }
\mnref{Eisner, J.A., Lane, B.F., Hillenbrand, L.A., Akeson, R.L., 
  Sargent, A.I., 2004, \apj, 613, 1049 }
\mnref{Felli, M., Massi, M., Staude, H.J., Reddmann, T., Eiroa, C., Hefele, H.,
 Neckel, T., Panagia, N., 1984, \aaa, 135, 261 }
\mnref{Folha, D.F.M., Emerson, J.P.\ 2001, \aap, 365, 90 }
\mnref{Follert, R., Linz, H., 
Stecklum, B., van Boekel, R., Henning, T., Feldt, M., Herbst, T.~M., 
\& Leinert, C., 2010, in press}
\mnref{Fuente, A., Martin-Pintado, J., Bachiller, R., Cernicharo, J.\ 1990,
  \aap, 237, 471 } 
\mnref{Gordon, M.A., 2003, \apj, 589, 953}
\mnref{Gordon, M.A., Holder, B.P., Jisonna, L.J., Jr., Jorgenson, R.A., 
  Strelnitski, V.S.\ 2001, \apj, 559, 402 }
\mnref{Hamann, F., Simon, M., 1986, ApJ, 311, 909}
\mnref{Hamann, F., Simon, M., 1988, ApJ, 339, 1078}
\mnref{Hamann, F., Persson, S.E., 1989, \apjs, 71, 931 }
\mnref{Hamann, F., Persson, S.E., 1992, \apjs, 82, 285 }
\mnref{Hamann, F., Depoy, D.L., Johansson, S., Elias, J., 1994, \apj, 422, 626 }
\mnref{Hanson, M.M., Howarth, I.D., Conti, P.S., 1997, \apj, 489, 698 }
\mnref{Hanson, M.M., Kudritzki, R.-P., Kenworthy, M.A., Puls, J., 
  Tokunaga, A.T.\ 2005, \apjs, 161, 154 }
\mnref{Harmanec, P., 1988, Bull.\ Astron.\ Inst.\ Czechoslovakia, 39, 329}
\mnref{Hern{\'a}ndez, J., Calvet, N., Brice{\~n}o, C., Hartmann, L., 
   Berlind, P., 2004, \aj, 127, 1682 }
\mnref{Hinkle, K., Wallace, L., Livingston, W., 1995, PASP, 107, 1042}
\mnref{Hippelein, H., Muench, G., 1981, \aaa, 99, 248} 
\mnref{Hoare, M.G., 2006, ApJ, 649, 856 }
\mnref{Hofmann, K.-H., Balega, Y., Ikhsanov, N.R., Miroshnichenko, A.S., 
   Weigelt, G., 2002, \aap, 395, 891 }
\mnref{Hoffmeister, V.H., Chini, R., Scheyda, C.M., Schulze, D., 
  Watermann, R., N{\"u}rnberger, D., Vogt, N.\ 2008, \apj, 686, 310 }
\mnref{Hollenbach, D., Johnstone, D., Lizano, S., Shu, F., 1994, 
  \apj, 428, 654}
\mnref{Hosokawa T., Omukai K., 2009, ApJ, 691, 823 }
\mnref{Johansson, S., 1978, Phys. Scr. 18, 217}
\mnref{Johansson, S., Letokhov, V.S., 2007, NewAR, 51, 443}
\mnref{Kahn, F.D., 1974, A\&A, 37, 149}
\mnref{Kelly, D.M., Rieke, G.H., Campbell, B., 1994, \apj, 425, 231 }
\mnref{Kleinmann, D.E., \& Wright, E.L., 1973, \apj, 185, L131 }
\mnref{Kraus, M., Kr{\"u}gel, E., Thum, C.,  Geballe, T.R., 2000, 
    \aap, 362, 158 }
\mnref{Kraus S., et al., 2010, Natur, 466, 339 }
\mnref{Krumholz, M.R., Klein, R.I., McKee, C.F., Offner, S.S.R., 
   \& Cunningham, A.J., 2009, Science, 323, 754 }
\mnref{Kuiper, R., Klahr, H., Beuther, H., Henning, T., 2010, ApJ, 722, 1556 }
\mnref{Kurosawa R., Romanova M.M., Harries T.J., 2011, MNRAS, 416, 2623 }
\mnref{Lada, C.J., 1976, \apjs, 32, 603 }
\mnref{Laor, A., 2006, ApJ, 643, 112}
\mnref{Lenorzer, A., de Koter, A., Waters, L.B.F.M., 2002, \aaa, 386, L5 }
\mnref{Linz, H. et al.\,2009, \aaa, 505, 655}
\mnref{Lumsden, S.L., Puxley, P.J., 1996, MNRAS, 281, 493}
\mnref{Malbet, F. et al., 2007, \aap, 464, 43 }

\mnref{McKee, C.F., Tan, J.C., 2003, \apj, 585, 850 }
\mnref{Mel\'{e}ndez, J., Barbuy, B., 1999, ApJS, 124, 527}
\mnref{Meyer, J.M., Nordsieck, K.H.,  Hoffman, J.L., 2002, \aj, 123, 1639 }
\mnref{Montesinos, B., Eiroa, C., Mora, A., Mer{\'{\i}}n, B.\ 2009, 
  \aap, 495, 901 }
\mnref{Murdoch, K.A., Drew, J.E.\ 1994, in 
  The, P.S., Perez, M.R., Van den Heuvel, E.P.J., eds, ASP Conf. Series 62,
``The Nature and Evolutionary Status of Herbig Ae/Be Stars'', 
 p. 377}
\mnref{Muzerolle, J., Calvet, N.,  Hartmann, L.\ 2001, \apj, 550, 944 }
\mnref{Oudmaijer, R.D., 1998, A\&AS, 129, 541}
\mnref{Oudmaijer, R.D., Busfield, G., Drew, J.E., 1997, \mnras, 291, 797 }
\mnref{Oudmaijer, R.D., Drew, J.E., 1999, \mnras, 305, 166 }
\mnref{Oudmaijer, R.D., Drew, J.E., Vink, J.S., 2005, \mnras, 364, 725 }
\mnref{Oudmaijer, R.D. et al.\ 2011, Astronomische Nachrichten, 332, 238} 
\mnref{Persson, S.E., Geballe, T.R., McGregor, P.J., Edwards, S., 
  Lonsdale, C.J., 1984, \apj, 286, 289 }
\mnref{Pontoppidan, K.M., Dullemond, C.P., Blake, G.A., Evans, N.J., II, 
Geers, V.C., Harvey, P.M., Spiesman, W., 2007, \apj, 656, 991 }
\mnref{Pontoppidan, K.M., Blake, G.A., Smette, A.\ 2011, \apj, 733, 84 }
\mnref{Porter, J.M., Drew, J.E., Lumsden, S.L. 1998, \aaa, 332, 999}
\mnref{Pringle, J.E., 1981, ARA\&A, 19, 137 }
\mnref{Prisinzano, L., Damiani, F., Micela, G., Sciortino, S., 2005, \aap, 430, 941 }
\mnref{Ramsauer J., Solanki S.K., Biemont E., 1995, \aas, 113, 71}
\mnref{Reipurth, B., Pedrosa, A., Lago, M.T.V.T., 1996, \aas, 120, 229 }
\mnref{Repolust T., Puls J., Hanson M.M., Kudritzki R.-P., Mokiem M.R., 
2005, A\&A, 440, 261 }
\mnref{Quirrenbach, A., Albrecht, S., Tubbs, R.N., 2006, 
  Kraus, M,  Miroshnichenko, A.S., eds,  ASP Conference
  Series, Vol. 355, ``Stars with the B[e] Phenomenon'', p.239 }
\mnref{Saito, H. et al., 2009, \aj, 137, 3149 }
\mnref{Schneider, N., Simon, R., Bontemps, S., Comer{\'o}n, F., Motte, F.,
  2007, \aap, 474, 873 }
\mnref{Schreyer, K., Henning, T., van der Tak, F.F.S., Boonman, A.M.S., van
  Dishoeck, E.F.,  2002, \aap, 394, 561 }
\mnref{Schreyer, K., Semenov, D., Henning, T., Forbrich, J., 2006, 
  \apj, 637, L129}
\mnref{Shu, F.H., Adams, F.C., Lizano, S., 1987, ARA\&A, 25, 23 }
\mnref{Sim, S.A., Drew, J.E., Long, K.S., 2005, \mnras, 363, 615 }
\mnref{Simon, M., Cassar, L., Felli, M., Fischer, J., Massi, M., 
  Sanders, D., 1984, \apj, 278, 170}
\mnref{Solf, J., Carsenty, U., 1982, \aaa, 113, 142 }
\mnref{Stahler, S.W., Palla, F., Ho, P.T.P., 2000, 
  Mannings, V., Boss, A.P., Russell, S.S., eds, Protostars and Planets IV, 
    p.327 }
\mnref{Storey, P.J., Hummer, D.G.\ 1995, \mnras, 272, 41 }

\mnref{Thum, C., Matthews, H.E., Martin-Pintado, J., Serabyn, E., 
  Planesas, P.,  Bachiller, R., 1994, \aaa, 283, 582 }
\mnref{Thronson, H.A., Jr., Loewenstein, R.F., 
  Stokes, G.M., 1979, \aj, 84, 1328 }
\mnref{Vakili, F. et al., 1998, \aap, 335, 261}
\mnref{van den Ancker, M.E., de Winter, D., Tjin A Djie, H.R.E., 1998, 
  \aap, 330, 145 }
\mnref{van den Ancker, M.E., Wesselius, P.R.,  Tielens, A.G.G.M., 2000, 
  \aap, 355, 194 }
\mnref{Vink, J.S., Drew, J.E., Harries, T.J., Oudmaijer, R.D., 2002, 
  \mnras, 337, 356 }
\mnref{Vink, J.S., Drew, J.E., Harries, T.J., Oudmaijer, R.D., 
  Unruh, Y., 2005, \mnras, 359, 1049 }
\mnref{Weigelt, G., et al.\ 2011, \aap, 527, A103 }
\mnref{Weintroub, J., Moran, J.M., Wilner, D.J., Young, K., Rao, R., 
    Shinnaga, H., 2008, \apj, 677, 1140 }
\mnref{Wheelwright, H.E., Oudmaijer, R.D., de Wit, W.J., Hoare, M.G., Lumsden,
  S.L., Urquhart, J.S., 2010, \mnras,  408, 1840}
\mnref{Willner, S.P. et al., 1982, \apj, 253, 174 }
\mnref{Wouterloot, J.G.A., Brand, J., 1989, A\&AS, 80, 149}
\mnref{Zorec, J., Arias, M.L., Cidale, L., Ringuelet, A.E., 2007, A\&A, 470, 239 }

\end{refs}

\end{document}